\documentclass[ reprint,twocolumn,groupedaddress]{revtex4-2}
\usepackage[T1]{fontenc}
\usepackage{amsmath}
\usepackage{amssymb}
\usepackage{graphicx}
\usepackage{dcolumn}
\usepackage{bm}

\begin{document}

\title{Channel-Differentiated Topological Charge Flips and Formation of Janus Bound States in the Continuum in Single-Layer Photonic Lattices}

\author{Sun-Goo Lee}
\email{sungooleee@gmail.com}
\affiliation{Department of Data Information and Physics, Kongju National University, Gongju, 32588, South Korea}
\affiliation{Institute of Application and Fusion for Light, Kongju National University, Cheonan, 31080, South Korea}

\author{Wook-Jae Lee}
\email{wookjaelee@kongju.ac.kr}
\affiliation{Department of Data Information and Physics, Kongju National University, Gongju, 32588, South Korea}
\affiliation{Institute of Application and Fusion for Light, Kongju National University, Cheonan, 31080, South Korea}
\date{\today}

\begin{abstract}
The topological charge of a bound state in the continuum (BIC) can flip when the BIC merges with other polarization singularities or, as recently shown, at band flips. Here, using a coupled-mode analysis, we derive an analytical expression for the topological charge of symmetry-protected BICs in one-dimensional single-layer photonic lattices. The expression captures both merging-induced and band-flip-induced charge flips. We find that, during a band flip, the topological charge of the BIC reverses with the formation of lines of circular polarization that cross at the $\Gamma$ point. In vertically symmetric lattices, each charge flip occurs simultaneously in the upward and downward radiation channels. Breaking vertical symmetry makes both kinds of charge flips channel-differentiated, creating a parameter interval in which the BIC carries opposite topological charges in the two channels and forms a Janus BIC. We show that Janus BICs also emerge in experimentally accessible one- and two-dimensional silicon-based photonic lattices. Our results link the topological charge of a BIC directly to the band structure and the channel-resolved radiation amplitudes and establish a basis for channel-selective control of polarization singularities.
\end{abstract}

\maketitle

Topological photonics has become a powerful framework for precisely engineering the radiative and polarization properties of light in open planar photonic structures, including one-dimensional (1D) gratings, two-dimensional (2D) photonic-crystal slabs, and metasurfaces~\cite{Zhen2014,Fan2002,Marinica2008,Hsu2013,Hsu2016,MKang2023}. Polarization singularities carrying quantized topological charges are ubiquitous in momentum-space polarization fields~\cite{Zhang2018,Zhou2018,Doeleman2018,Ye2020}, and their controlled manipulation has enabled a variety of intriguing topological optical phenomena, including bound states in the continuum (BICs) with ultrahigh $Q$ factors~\cite{Koshelev2018,Minkov2018,Kodigala2017,MSHwang2021,SGLee2021PRL}, chiral emission from circularly polarized states (C points)~\cite{Liu2019CP,Gorkunov2020,Overvig2021,Zhang2022,Chen2023}, one-sided radiation in unidirectional guided resonances (UGRs)~\cite{XYin2020,Yin2023,Su2026,SGLee2025PR}, and opposite topological charges of Janus BICs in the upward and downward radiation channels~\cite{Kang2025Janus,Zuo2025Janus,Ji2026Janus,Zhang2026Vec,Su2026Nano}. Although the total topological charge is conserved, individual polarization singularities can be created, displaced, split, and merged as the photonic lattice is continuously deformed~\cite{Yoda2020,Kang2021,Hu2022}.

Recent studies have explored the control of the topological charge of BICs by varying structural parameters. Switching the charge reconfigures the momentum-space polarization texture of the far-field radiation, providing a route to tunable structured-light generation with applications such as chiral emission and vortex-beam engineering. The charge of a BIC can change through interactions with other polarization singularities, as BICs or C points merge with it or split off from it~\cite{JJin2019,Kang2022,LiLPR2026,Wang2026Bilayer,Tu2026}. The charge can also flip at a band inversion, as the BIC exchanges its band character with a radiative partner~\cite{Bouteyre2026}, or under a shear deformation that breaks the in-plane mirror symmetries and drives the charge of an isolated BIC through zero~\cite{Li2026ChargeFree}. In these studies, the far-field coefficients near a BIC that determine its charge have generally been introduced as phenomenological parameters or extracted from numerical simulations.

In this Letter, we derive an analytical expression for the topological charge of symmetry-protected BICs in 1D single-layer photonic lattices using a coupled-mode model~\cite{Kazarinov1985,BandFlip,SGLee2023}. The expression relates the charge to the band structure and the channel-resolved radiation amplitudes, thereby capturing two kinds of charge flips. In one, off-$\Gamma$ BICs or C points merge with the symmetry-protected BIC at the $\Gamma$ point while the band gap remains open, producing a merging-induced charge flip. In the other, two lines of circular polarization that cross at the $\Gamma$ point are formed as the band gap closes, producing a band-flip-induced charge flip. When vertical symmetry is broken, both kinds of charge flips become channel-differentiated, with the upward and downward radiation channels flipping at different values of a lattice parameter. Between the two charge-flip points, the BIC carries opposite topological charges in the two channels, forming a Janus BIC. Full-wave calculations confirm the predictions and show that Janus BICs also emerge in experimentally realizable 1D and 2D lattices based on partially etched silicon slabs.

\begin{figure}[t!]
\centering
\includegraphics[width=8cm]{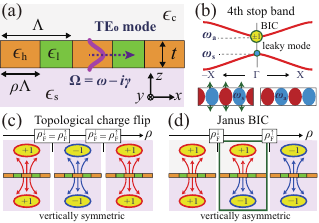}
\caption{\label{fig1} (a) Schematic of the 1D photonic lattice. (b) Fourth stop band of the TE$_0$ mode with a symmetry-protected BIC. (c) Simultaneous charge flips in a vertically symmetric lattice. (d) Channel-differentiated charge flips and a Janus BIC in a vertically asymmetric lattice.}
\end{figure}

Figure~\ref{fig1}(a) illustrates a representative single-layer 1D photonic lattice. It consists of alternating high- and low-dielectric-constant regions $\epsilon_{\mathrm h}$ and $\epsilon_{\mathrm l}$ with period $\Lambda$, thickness $t$, and duty cycle $\rho$, bounded by a cover $\epsilon_{\mathrm c}$ and a substrate $\epsilon_{\mathrm s}$. The grating supports TE-polarized guided modes because its average dielectric constant $\epsilon_{\mathrm{avg}}=\epsilon_{\mathrm l}+\rho\Delta\epsilon$, with $\Delta\epsilon=\epsilon_{\mathrm h}-\epsilon_{\mathrm l}$, exceeds those of the surrounding media. We study charge flips of symmetry-protected BICs at the edges of the fourth stop band of the TE$_0$ mode~\cite{SGLee2021PR,SGLee2024RiP,SGLee2026Nano}, shown in Fig.~\ref{fig1}(b). At this stop band, varying the duty cycle drives both merging-induced and band-flip-induced charge flips. The band-edge mode at $\omega_{\mathrm a}$ has an antisymmetric field distribution and is the BIC, carrying a topological charge of $+1$ or $-1$, whereas the mode at $\omega_{\mathrm s}$ is symmetric and leaky, radiating upward and downward. Figures~\ref{fig1}(c) and \ref{fig1}(d) schematically depict the charge flips in vertically symmetric and asymmetric lattices, respectively. The vertical mirror symmetry forces the two channels to flip at the same duty cycle, $\rho_{\mathrm F}^{\uparrow}=\rho_{\mathrm F}^{\downarrow}$, whereas broken vertical symmetry makes the two channels inequivalent~\cite{Glowadzka2021,Bai2025} and separates their charge-flip points, $\rho_{\mathrm F}^{\uparrow}\neq\rho_{\mathrm F}^{\downarrow}$, opening a Janus window~\cite{Kang2025Janus}.

To elucidate the charge flips and Janus BICs, we start from the vector wave equation
\begin{equation}\label{eq:wave}
\nabla\times\nabla\times\mathbf E(x,y,z)-\epsilon(x,z)k_0^2\,\mathbf E(x,y,z)=0,
\end{equation}
where $k_0$ is the free-space wavenumber. The dielectric function of the $C_2^z$-symmetric lattice is expanded as $\epsilon(x,z)=\epsilon_0(z)+\sum_{n\ge1}\epsilon_n(z)\cos(nKx)$, with $K=2\pi/\Lambda$ and, inside the grating layer, $\epsilon_0=\epsilon_{\mathrm{avg}}$ and $\epsilon_n=(2\Delta\epsilon/n\pi)\sin(n\pi\rho)$. Near the third-order $\Gamma$ point, we extend the Kazarinov--Henry model~\cite{Kazarinov1985,BandFlip,SGLee2023} to in-plane wave vectors with both $k_x$ and $k_y$ components. Retaining Fourier components up to $\epsilon_4$, we approximate the electric field as
\begin{multline}\label{eq:field}
\mathbf E=\Big[\big(A\hat{\mathbf e}_+e^{i2Kx}+B\hat{\mathbf e}_-e^{-i2Kx}\big)\phi+\mathbf E_{\mathrm{rad}}\\
+\sum_{m=1,3}\big(\mathbf E_{+m}e^{imKx}+\mathbf E_{-m}e^{-imKx}\big)\Big]e^{i(k_xx+k_yy)}.
\end{multline}
Here, $A$ and $B$ are the amplitudes of the counterpropagating TE$_0$ guided waves, $\hat{\mathbf e}_{\pm}$ are their polarization vectors, $\phi$ is the transverse profile of the TE$_0$ mode of the unmodulated reference slab $\epsilon_0(z)$, and $\mathbf E_{\mathrm{rad}}$ and $\mathbf E_{\pm m}$ are the $z$-dependent amplitudes of the zeroth-order and intermediate harmonics. Substituting the truncated dielectric function and Eq.~\eqref{eq:field} into Eq.~\eqref{eq:wave} and projecting the $\pm2K$ components onto $\phi$ gives the coupled-mode equation at the in-plane wave vector $\mathbf k=(k_x,k_y)$,
\begin{equation}\label{eq:KH}
Kn_{\mathrm g}(\Omega_0-\Omega)\begin{pmatrix}A\\B\end{pmatrix}=\begin{pmatrix}h_{AA}-k_x&h_{AB}\\h_{BA}&h_{BB}+k_x\end{pmatrix}\begin{pmatrix}A\\B\end{pmatrix},
\end{equation}
where $\Omega=\omega-i\gamma$ is the complex frequency in units of $2\pi c/\Lambda$, $\Omega_0$ is the Bragg frequency of the reference slab, $n_{\mathrm g}$ is the group index of the guided wave, and $\pm k_x$ is the detuning of each guided wave from the Bragg condition. The coupling coefficients, which depend on $\mathbf k$ and satisfy $h_{AA}=h_{BB}$ and $h_{AB}=h_{BA}$, are given in the Supplemental Material~\cite{SM}. For $k_y=0$, the eigenvalues of Eq.~\eqref{eq:KH} near the $\Gamma$ point give the dispersion relation along $k_x$~\cite{BandFlip},
\begin{equation}\label{eq:dispersion}
\Omega(k_x)=\Omega_0-\left(ih_{\mathrm r}\pm\sqrt{k_x^2+(h_{\mathrm g}+ih_{\mathrm r})^2}\right)/(Kn_{\mathrm g}),
\end{equation}
where $h_{\mathrm g}=h_{AB}-h_{AA}$ denotes the nonradiative Bragg coupling and $h_{\mathrm r}=-ih_{AA}$ represents the radiative coupling through the zeroth-order channel. Equation~\eqref{eq:dispersion} gives two band edges at $k_x=0$, the BIC ($A=-B$) at $\Omega_{\mathrm a}=\Omega_0+h_{\mathrm g}/(Kn_{\mathrm g})$ and the leaky state ($A=B$) at $\Omega_{\mathrm s}=\Omega_0-(h_{\mathrm g}+2ih_{\mathrm r})/(Kn_{\mathrm g})$. The band gap $2|h_{\mathrm g}-\mathrm{Im}\,h_{\mathrm r}|/(Kn_{\mathrm g})$ closes at $G_{\mathrm{band}}\equiv h_{\mathrm g}-\mathrm{Im}\,h_{\mathrm r}=0$, across which the BIC moves to the opposite band edge in a band flip~\cite{BandFlip}.

The field radiated into the cover ($\sigma=\,\uparrow$) and the substrate ($\sigma=\,\downarrow$) is
\begin{equation}\label{eq:Erad}
\mathbf E_{\mathrm{rad}}^{\sigma}(\mathbf k)=-\frac{k_0^2}{2}\mathcal K^{\sigma}\!\int u^{\sigma}(z)\,\mathbf S(z)\,dz,
\end{equation}
where $\mathcal K^{\sigma}$ and $u^{\sigma}$ are the coupling constant and the outgoing radiation profile of channel $\sigma$~\cite{SM}. The zeroth-order source is $\mathbf S=A\mathbf S_++B\mathbf S_-$, where $\mathbf S_\pm$ are produced by the $\pm2K$ guided waves of unit amplitude~\cite{SM}, and $(A,B)$ is the eigenvector of Eq.~\eqref{eq:KH}. The topological charge of channel $\sigma$ is the winding number of the polarization orientation of $\mathbf E_{\mathrm{rad}}^{\sigma}$ along a closed loop $C$ encircling the $\Gamma$ point~\cite{Zhen2014},
\begin{equation}\label{eq:winding}
q^{\sigma}=\frac{1}{2\pi}\oint_C d\mathbf k\cdot\nabla_{\mathbf k}\psi^{\sigma}(\mathbf k),\quad \psi^{\sigma}=\frac12\arg\big(S_1^{\sigma}+iS_2^{\sigma}\big),
\end{equation}
where $S_0=|E_x|^2+|E_y|^2$, $S_1=|E_x|^2-|E_y|^2$, and $S_2=2\,\mathrm{Re}(E_x^*E_y)$ are the Stokes parameters of $\mathbf E_{\mathrm{rad}}^{\sigma}$. With Eq.~\eqref{eq:Erad} and the eigenvectors of Eq.~\eqref{eq:KH}, Eq.~\eqref{eq:winding} gives the topological charge without further approximation~\cite{SM}.

We now derive an analytical expression for the BIC charge $q^{\sigma}$ from the far-field radiation near the $\Gamma$ point. On the $k_x$ axis, the source $\mathbf S=A\mathbf S_++B\mathbf S_-$ is $y$ polarized and takes the form $\mathbf S=(A+B)\tilde S_y\hat{\mathbf y}$ with $\tilde S_y=(AS_{+,y}+BS_{-,y})/(A+B)$. On the $k_y$ axis, where $A=-B$, the tilt of $\hat{\mathbf e}_{\pm}$ gives the two guided waves $x$-polarized sources of opposite sign, and the source takes the form $\mathbf S=-(A-B)(k_y/2K)\tilde S_x\hat{\mathbf x}$, where $\tilde S_x$ includes the TM responses induced by the tilt~\cite{SM}. Thus $A+B$ determines the $y$-polarized radiation and $A-B$ the $x$-polarized radiation. The detuning $\pm k_x$ in Eq.~\eqref{eq:KH} makes the amplitudes of the two guided waves unequal, giving $A+B=ak_x/(h_{\mathrm g}+ih_{\mathrm r})$ and $A-B=2a$ on the BIC band, where $A=-B=a$ at the $\Gamma$ point. Inserting the source and the amplitudes into Eq.~\eqref{eq:Erad} gives $\mathbf E_{\mathrm{rad}}^{\sigma}\simeq(c^{\sigma}k_y,\,d^{\sigma}k_x)$ with
\begin{equation}\label{eq:cd}
c^{\sigma}=\frac{k_0^2}{2K}\,\mathcal K^{\sigma}a\,F_x^{\sigma},\qquad d^{\sigma}=-\frac{k_0^2}{2}\,\frac{\mathcal K^{\sigma}a}{h_{\mathrm g}+ih_{\mathrm r}}\,F_y^{\sigma},
\end{equation}
where $F_x^{\sigma}$ and $F_y^{\sigma}$ are the $\mathbf k\to0$ limits of the radiation factors of the $x$- and $y$-polarized sources into channel $\sigma$,
\begin{equation}\label{eq:F}
F_x^{\sigma}(k_y)=\int u^{\sigma}\tilde S_x\,dz,\qquad F_y^{\sigma}(k_x)=\int u^{\sigma}\tilde S_y\,dz.
\end{equation}

Substituting $\mathbf E_{\mathrm{rad}}^{\sigma}\simeq(c^{\sigma}k_y,\,d^{\sigma}k_x)$ into Eq.~\eqref{eq:winding} gives $q^{\sigma}=-\mathrm{sgn}\,\mathrm{Re}(c^{\sigma*}d^{\sigma})$~\cite{SM}. Because the far field is linear in $\mathbf k$, $|q^{\sigma}|=1$ whenever $\mathrm{Re}(c^{\sigma*}d^{\sigma})\neq0$. Equation~\eqref{eq:cd} therefore gives $q^{\sigma}=\mathrm{sgn}\,\mathrm{Re}\big[F_x^{\sigma*}F_y^{\sigma}/(h_{\mathrm g}+ih_{\mathrm r})\big]$. Writing $h_{\mathrm g}+ih_{\mathrm r}=G_{\mathrm{band}}+i\,\mathrm{Re}\,h_{\mathrm r}$, we obtain
\begin{equation}\label{eq:qa}
q^{\sigma}=\mathrm{sgn}\big[G_{\mathrm{band}}\,\mathrm{Re}\big(F_x^{\sigma*}F_y^{\sigma}\big)+\mathrm{Re}\,h_{\mathrm r}\,\mathrm{Im}\big(F_x^{\sigma*}F_y^{\sigma}\big)\big].
\end{equation}

\begin{figure}[t]
\centering
\includegraphics[width=\columnwidth]{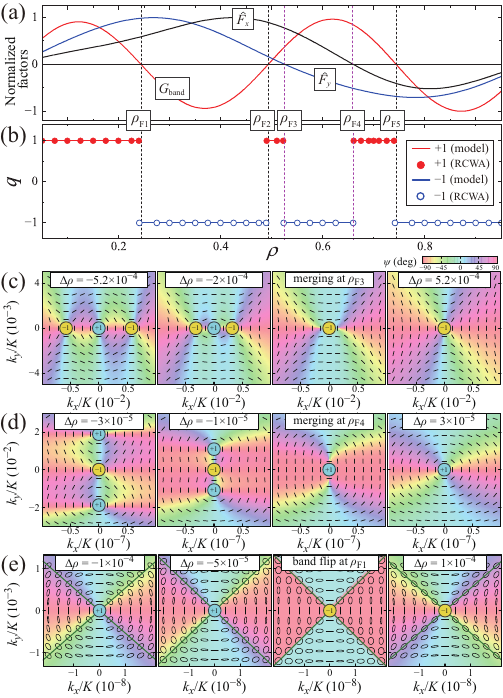}
\caption{Charge flips in the vertically symmetric lattice ($\epsilon_{\mathrm c}=\epsilon_{\mathrm s}=2.1316$, $\epsilon_{\mathrm{avg}}=12$, $\Delta\epsilon=1$, $t=0.5\Lambda$), calculated from the model. (a) Normalized $G_{\mathrm{band}}$, $\hat F_y$, and $\hat F_x$. (b) Topological charges, with RCWA results for comparison. (c),(d) Far-field polarization orientation as the off-$\Gamma$ BIC pair on the $k_x$ axis (c) and the $k_y$ axis (d) merges with the symmetry-protected BIC at $\rho_{\mathrm{F3}}$ and $\rho_{\mathrm{F4}}$, respectively. (e) Far-field polarization ellipses near the band flip at $\rho_{\mathrm{F1}}$. Green lines mark the directions in which $|E_x|=|E_y|$. Colors in (c)--(e) show the polarization orientation angle $\psi$. $\Delta\rho$ is measured from the charge-flip point.}
\label{fig2}
\end{figure}

Equation~\eqref{eq:qa} reveals that the charge of channel $\sigma$ depends on the real part $\mathrm{Re}(F_x^{\sigma*}F_y^{\sigma})$ and the imaginary part $\mathrm{Im}(F_x^{\sigma*}F_y^{\sigma})$ of the product of the radiation factors, weighted by the band gap $G_{\mathrm{band}}$ and the radiative coupling $\mathrm{Re}\,h_{\mathrm r}$. For a vertically symmetric lattice, $\phi$ and the responses of the $\pm K$ and $\pm3K$ harmonics are real and even in $z$. The two channels then have identical radiation factors, and, to leading order in the modulation, $F_x^{\sigma}$ and $F_y^{\sigma}$ have a common phase, $F_j^{\sigma}\propto\hat F_j$ with real $\hat F_j$~\cite{SM}. The imaginary part of $F_x^{\sigma*}F_y^{\sigma}$ thus vanishes, and Eq.~\eqref{eq:qa} reduces to
\begin{equation}\label{eq:q}
q^{\uparrow}=q^{\downarrow}=\mathrm{sgn}\big[\hat F_x\hat F_yG_{\mathrm{band}}\big].
\end{equation}
In a vertically symmetric lattice, therefore, the topological charge flips when one of the three real factors $\hat F_x$, $\hat F_y$, and $G_{\mathrm{band}}$ changes sign. Figures~\ref{fig2}(a) and \ref{fig2}(b) show the three factors as functions of $\rho$ and the five resulting charge flips at $\rho_{\mathrm{F1}}$--$\rho_{\mathrm{F5}}$, all reproduced by rigorous coupled-wave analysis (RCWA)~\cite{Moharam1995,Li1996}. At $\rho_{\mathrm{F3}}$ and $\rho_{\mathrm{F4}}$, the charge flips when $\hat F_y$ and $\hat F_x$ vanish, respectively, while the band gap remains open. Because the far field is purely $y$-polarized along $k_x$ and purely $x$-polarized along $k_y$, a zero of $F_y^{\sigma}(k_x)$ or $F_x^{\sigma}(k_y)$ corresponds to an off-$\Gamma$ BIC~\cite{SGLee2026Nano}, and $C_2^z$ symmetry places such BICs in pairs at $\pm k_x$ or $\pm k_y$~\cite{SM}. As $\rho$ varies, each pair approaches $\Gamma$ and merges with the symmetry-protected BIC when the corresponding factor at $\Gamma$ vanishes, reversing its charge~\cite{Zhen2014,SGLee2023}, as shown in Figs.~\ref{fig2}(c) and \ref{fig2}(d). At $\rho_{\mathrm{F1}}$, $\rho_{\mathrm{F2}}$, and $\rho_{\mathrm{F5}}$, the charge instead flips at a band flip, where $G_{\mathrm{band}}$ vanishes~\cite{Bouteyre2026}. Near the $\Gamma$-point BIC, the model gives the $y$-polarized far field on the $k_x$ axis, $E_y^{\sigma}=-(k_0^2/2)\mathcal K^{\sigma}(A+B)F_y^{\sigma}$ with $A+B=ak_x/(G_{\mathrm{band}}+i\,\mathrm{Re}\,h_{\mathrm r})$. At the band flip, $G_{\mathrm{band}}=0$ and $A+B=-iak_x/\mathrm{Re}\,h_{\mathrm r}$, whereas $A-B$, which sets $E_x^{\sigma}$, stays at $2a$. Because $F_x^{\sigma}$ and $F_y^{\sigma}$ share a common phase, $E_y^{\sigma}$ is then $90^\circ$ out of phase with $E_x^{\sigma}$, and the far field forms lines of circular polarization through $\Gamma$ along the directions in which $|E_x^{\sigma}|=|E_y^{\sigma}|$~\cite{SM}. Figure~\ref{fig2}(e) shows that the polarization along these lines becomes increasingly circular as $\rho$ approaches $\rho_{\mathrm{F1}}$, where the charge flips from $+1$ to $-1$.

\begin{figure}[t]
\centering
\includegraphics[width=\columnwidth]{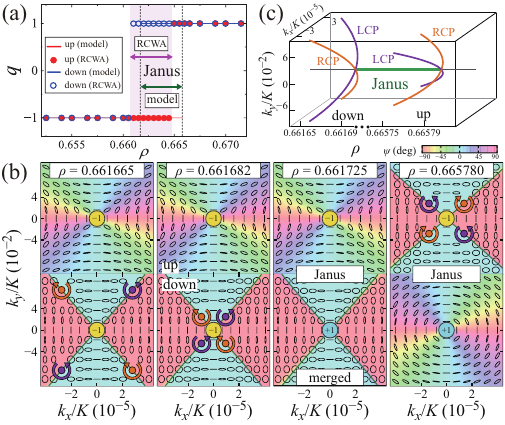}
\caption{Janus BICs in the vertically asymmetric lattice ($\epsilon_{\mathrm c}=1$, $\epsilon_{\mathrm s}=2.1316$, $\epsilon_{\mathrm{avg}}=12$, $\Delta\epsilon=1$, $t=0.5\Lambda$), calculated from the model. (a) Channel charges near $\rho_{\mathrm{F4}}$, with RCWA results. (b) Far-field polarization maps for three values of $\rho$ around the downward-channel flip and one just below the upward-channel flip. Green lines mark the directions in which $|E_x^{\sigma}|=|E_y^{\sigma}|$ near each channel flip. Colors show the polarization orientation angle $\psi$. (c) C-point trajectories.}
\label{fig3}
\end{figure}

In a vertically asymmetric lattice, $F_x^{\sigma}$ and $F_y^{\sigma}$ become complex~\cite{SM}, allowing the charge to flip without either factor vanishing. Near $\rho_{\mathrm{F3}}$ and $\rho_{\mathrm{F4}}$, where the symmetric lattice flips its charge at $\hat F_y=0$ or $\hat F_x=0$, the charge of channel $\sigma$ flips where $G_{\mathrm{band}}\mathrm{Re}(F_x^{\sigma*}F_y^{\sigma})+\mathrm{Re}\,h_{\mathrm r}\mathrm{Im}(F_x^{\sigma*}F_y^{\sigma})$ in Eq.~\eqref{eq:qa} vanishes. The off-$\Gamma$ BICs, which require $F_y^{\sigma}(k_x)=0$ or $F_x^{\sigma}(k_y)=0$, also disappear, because the real and imaginary parts of $F_y^{\sigma}(k_x)$ or $F_x^{\sigma}(k_y)$ do not vanish at the same point on the corresponding axis. Each off-$\Gamma$ BIC splits into a right-handed circularly polarized (RCP) and a left-handed circularly polarized (LCP) C point, each carrying a half-integer charge~\cite{Liu2019CP,Yoda2020}. The C points of channel $\sigma$ reach $\Gamma$ at its charge flip. Because the two channels radiate into different claddings, their radiation factors have different phases, thereby shifting the flip to different values of $\rho$ for the two channels~\cite{XYin2020,SGLee2025PR} and opening Janus windows between them. Figure~\ref{fig3}(a) shows the window near $\rho_{\mathrm{F4}}$, in which the downward channel flips from $q^{\downarrow}=-1$ to $+1$ at $\rho=0.66168$ in the model, while the upward channel keeps $q^{\uparrow}=-1$ until $\rho=0.66579$. RCWA gives a window of nearly the same width between $\rho=0.66094$ and $0.66489$. As shown in Fig.~\ref{fig3}(b), four C points of the downward channel approach $\Gamma$ and merge with the BIC, reversing its charge to $q^{\downarrow}=+1$, whereas the upward field remains a saddle with $q^{\uparrow}=-1$, so that the BIC becomes a Janus BIC. Near each channel flip, $\mathrm{Re}(c^{\sigma*}d^{\sigma})\approx0$, so that the far field of channel $\sigma$ is nearly circularly polarized along the directions in which $|E_x^{\sigma}|=|E_y^{\sigma}|$, and the C points of channel $\sigma$ move toward $\Gamma$ along these directions. Figure~\ref{fig3}(c) traces the C points, which reach $\Gamma$ at the two channel flips, and shows the resulting Janus window.

\begin{figure}[t]
\centering
\includegraphics[width=\columnwidth]{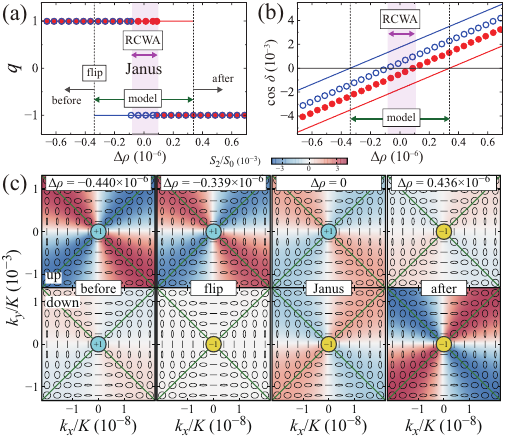}
\caption{Janus BIC near the first band flip of the lattice of Fig.~\ref{fig3}, calculated from the model. (a) Channel charges and (b) $\cos\delta^{\sigma}$ in the first quadrant of the far field, with RCWA results for comparison as in Fig.~\ref{fig3}(a). $\Delta\rho$ is measured from the center of each window, located $5.6\times10^{-8}$ above the band-flip point in the model and $4.35\times10^{-6}$ above the corresponding point in RCWA. (c) Far-field polarization maps of the upward and downward channels, overlaid on $S_2/S_0$. Green lines mark the directions in which $|E_x^{\sigma}|=|E_y^{\sigma}|$.}
\label{fig4}
\end{figure}

Equation~\eqref{eq:qa} also reveals that, in a vertically asymmetric lattice, the band-flip-induced charge flips are displaced from the band flip. With $\theta^{\sigma}=\arg(F_x^{\sigma*}F_y^{\sigma})$, the phase of $F_y^{\sigma}$ relative to $F_x^{\sigma}$, the charge of channel $\sigma$ flips at $G_{\mathrm{band}}=-\tan\theta^{\sigma}\,\mathrm{Re}\,h_{\mathrm r}$ for $\theta^{\sigma}\neq0$. At the band flip itself, Eq.~\eqref{eq:qa} reduces to $\mathrm{sgn}[\mathrm{Re}\,h_{\mathrm r}\,\mathrm{Im}(F_x^{\sigma*}F_y^{\sigma})]$, so the closing of the gap does not by itself flip the charge. Because $\theta^{\uparrow}\neq\theta^{\downarrow}$, the two channels flip at different values of $\rho$, creating a Janus window between them. To first order in $\mathrm{Re}\,h_{\mathrm r}$, its width is~\cite{SM}
\begin{equation}\label{eq:width}
\Delta\rho_{\mathrm J}\simeq\frac{|\tan\theta^{\uparrow}-\tan\theta^{\downarrow}|\,k_{\mathrm{EP}}}{|dG_{\mathrm{band}}/d\rho|},
\end{equation}
where $k_{\mathrm{EP}}=|\mathrm{Re}\,h_{\mathrm r}|$ is the exceptional-point momentum at which the two complex eigenfrequencies of Eq.~\eqref{eq:dispersion} become fully degenerate with the band-gap closure~\cite{BandFlip}, and $dG_{\mathrm{band}}/d\rho$ is evaluated there. At the band flip, $A+B=-iak_x/\mathrm{Re}\,h_{\mathrm r}$, so $k_{\mathrm{EP}}$ sets the momentum scale for hybridization between the BIC and leaky bands.

For the lattice of Fig.~\ref{fig3}, the model predicts three band flips over $0<\rho<1$, near which extremely narrow Janus windows of order $10^{-7}$ in $\rho$ open. The weak modulation of the lattice keeps the radiative coupling, and hence $k_{\mathrm{EP}}$, small. The flat upper and lower interfaces of the lattice also keep $|\tan\theta^{\uparrow}-\tan\theta^{\downarrow}|$ small. Equation~\eqref{eq:width} therefore yields such narrow windows. Figure~\ref{fig4}(a) shows that, near the first band flip, $\rho_{\mathrm{BF1}}=0.24366$ ($0.24173$ in RCWA), the downward charge flips before the upward charge, opening a window $6.8\times10^{-7}$ wide in the model and $1.8\times10^{-7}$ wide in RCWA. Figure~\ref{fig4}(b) shows $\cos\delta^{\sigma}=\mathrm{Re}(E_x^{\sigma*}E_y^{\sigma})/|E_x^{\sigma}E_y^{\sigma}|$, which gives the charge as $q^{\sigma}=-\mathrm{sgn}\cos\delta^{\sigma}$. Since $\cos\delta^{\sigma}=-(G_{\mathrm{band}}\cos\theta^{\sigma}+\mathrm{Re}\,h_{\mathrm r}\sin\theta^{\sigma})/[G_{\mathrm{band}}^2+(\mathrm{Re}\,h_{\mathrm r})^2]^{1/2}$, the curves of the two channels are parallel in both the model and RCWA, reflecting their common $G_{\mathrm{band}}$ dependence. The curves are offset because $\theta^{\uparrow}\neq\theta^{\downarrow}$, so that each crosses zero at the charge flip of its channel. The far-field polarization in Fig.~\ref{fig4}(c), overlaid on $S_2/S_0$, shows the corresponding reversal of the sign of $S_2$ in each quadrant. When $\cos\delta^{\downarrow}$ in Fig.~\ref{fig4}(b) crosses zero, $E_y^{\downarrow}$ is $90^\circ$ out of phase with $E_x^{\downarrow}$, and the downward far field forms lines of circular polarization through $\Gamma$ along the directions in which $|E_x^{\downarrow}|=|E_y^{\downarrow}|$. The downward charge thus reverses with the formation of these lines, as shown in Fig.~\ref{fig4}(c). Inside the window, the two channels carry opposite charges. Table~\ref{tab1} in the End Matter lists the widths $\Delta\rho_{\mathrm J}$ of the Janus windows near the three band flips for $\Delta\epsilon=1$, 2, 4, and 6. The windows widen by up to nearly three orders of magnitude as $\Delta\epsilon$ increases from 1 to 6 because both $k_{\mathrm{EP}}$ and $|\tan\theta^{\uparrow}-\tan\theta^{\downarrow}|$ grow with the modulation.

The End Matter extends the analysis to the second stop band (Fig.~\ref{fig5}), where merging-induced charge flips are driven by thickness variation~\cite{Zhen2014,SGLee2023} and band-flip-induced charge flips by $\rho$ variation~\cite{Bouteyre2026}, and to realistic 1D and 2D lattices (Figs.~\ref{fig6} and \ref{fig7}) based on partially etched silicon slabs, known as zero-contrast gratings (ZCGs)~\cite{Magnusson2014,ShokoohSaremi2014}, whose structural vertical asymmetry brings the range of lattice parameters hosting Janus BICs to an experimentally accessible level. Independent finite-element method (FEM) simulations also confirm the predicted charge flips and Janus BICs in Figs.~\ref{fig2}--\ref{fig4} and in the End Matter~\cite{SM}.

In summary, we have derived an analytical expression for the topological charge of symmetry-protected BICs at the fourth stop band of 1D photonic lattices. The expression captures both merging-induced charge flips, which occur when off-$\Gamma$ BICs or C points merge with the symmetry-protected BIC while the gap remains open, and band-flip-induced charge flips, which occur when two lines of circular polarization that cross at the $\Gamma$ point are formed as the band gap closes. In a vertically asymmetric lattice, the two radiation channels flip at different values of a lattice parameter, and between the two charge-flip points the BIC carries opposite topological charges in the upward and downward channels, forming a Janus BIC. We have extended the coupled-mode analysis to 2D lattices~\cite{SM}, and independent RCWA and FEM simulations confirm the predicted charge flips and Janus BICs. We have also shown that Janus BICs emerge in experimentally accessible 1D and 2D silicon ZCGs. Our results connect the topological charge of BICs to the coupled-mode description of photonic lattices and establish a basis for channel-selective control of polarization singularities.

\section*{End Matter}
\setcounter{equation}{0}
\renewcommand{\theequation}{A\arabic{equation}}

\begin{figure}[t]
\centering
\includegraphics[width=\columnwidth]{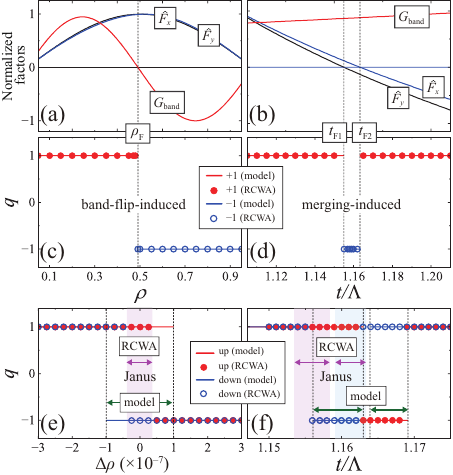}
\caption{Charge flips and Janus BICs at the second stop band of the TE$_0$ mode under duty-cycle variation at $t=0.5\Lambda$ (left column) and thickness variation at $\rho=0.30$ (right column), calculated from the model. (a),(b) Normalized $G_{\mathrm{band}}$, $\hat F_x$, and $\hat F_y$ for duty-cycle and thickness variation, respectively. (c),(d) Topological charges in the symmetric lattice of Fig.~\ref{fig2}, showing band-flip-induced and merging-induced charge flips, respectively. (e),(f) Channel charges in the asymmetric lattice of Fig.~\ref{fig3}, showing Janus windows. RCWA results are included in (c)--(f) for comparison. In (e), $\Delta\rho$ is measured from the center of each window.}
\label{fig5}
\end{figure}

\textit{Second stop band}---At the second stop band, the counterpropagating TE$_0$ guided waves at $\pm K$ radiate through $\epsilon_1$. The sources, the radiation factors, and the band coefficients $h_{\mathrm g}$ and $h_{\mathrm r}$ are given in the Supplemental Material~\cite{SM}. With these $h_{\mathrm g}$, $h_{\mathrm r}$, and $F_j^{\sigma}$, Eqs.~\eqref{eq:qa} and \eqref{eq:q} apply unchanged. In the symmetric lattice of Fig.~\ref{fig2} at $t=0.5\Lambda$, $\hat F_x$ and $\hat F_y$ follow $\epsilon_1\propto\sin\pi\rho$ and have no zero for $0<\rho<1$, so that the charge flips once at the band flip, at $\rho=0.49156$ in the model and $\rho=0.48905$ in RCWA, as shown in Figs.~\ref{fig5}(a) and \ref{fig5}(c). In the asymmetric lattice of Fig.~\ref{fig3}, the charge of channel $\sigma$ flips at $G_{\mathrm{band}}=-\tan\theta^{\sigma}\,\mathrm{Re}\,h_{\mathrm r}$. Because $\theta^{\uparrow}\neq\theta^{\downarrow}$, the two flips are displaced from the band flip by different amounts and open a Janus window, which Fig.~\ref{fig5}(e) shows to be $2.0\times10^{-7}$ wide in the model and $8.2\times10^{-8}$ wide in RCWA. Varying $t$ at $\rho=0.30$ instead produces merging-induced flips. As shown in Fig.~\ref{fig5}(b), $G_{\mathrm{band}}$ keeps its sign, whereas $\hat F_y$ and $\hat F_x$ pass through zero at $t=1.1550\Lambda$ and $1.1633\Lambda$, and Fig.~\ref{fig5}(d) shows the charge flipping at both thicknesses. RCWA places the two flips at $t=1.1545\Lambda$ and $1.1624\Lambda$. In the asymmetric lattice, the two channels flip at different thicknesses, and Fig.~\ref{fig5}(f) shows the resulting two Janus windows, with the same charge order and nearly the same widths in the model ($0.0069\Lambda$ and $0.0052\Lambda$) and RCWA ($0.0068\Lambda$ and $0.0056\Lambda$).

\begin{table}[t]
\caption{Widths $\Delta\rho_{\mathrm J1}$, $\Delta\rho_{\mathrm J2}$, and $\Delta\rho_{\mathrm J3}$ of the Janus windows near the three band flips, $\rho_{\mathrm{BF1}}<\rho_{\mathrm{BF2}}<\rho_{\mathrm{BF3}}$, of the lattice of Fig.~\ref{fig3} for $\Delta\epsilon=1$, 2, 4, and 6, calculated by RCWA.}
\label{tab1}
\begin{ruledtabular}
\begin{tabular}{cccc}
$\Delta\epsilon$ & $\Delta\rho_{\mathrm J1}$ & $\Delta\rho_{\mathrm J2}$ & $\Delta\rho_{\mathrm J3}$\\
\hline
1 & $1.8\times10^{-7}$ & $5.0\times10^{-8}$ & $2.6\times10^{-7}$\\
2 & $1.1\times10^{-6}$ & $3.8\times10^{-7}$ & $2.7\times10^{-6}$\\
4 & $5.5\times10^{-6}$ & $2.9\times10^{-6}$ & $5.4\times10^{-5}$\\
6 & $8.6\times10^{-6}$ & $3.8\times10^{-5}$ & $5.8\times10^{-5}$\\
\end{tabular}
\end{ruledtabular}
\end{table}

\begin{figure}[b]
\centering
\includegraphics[width=\columnwidth]{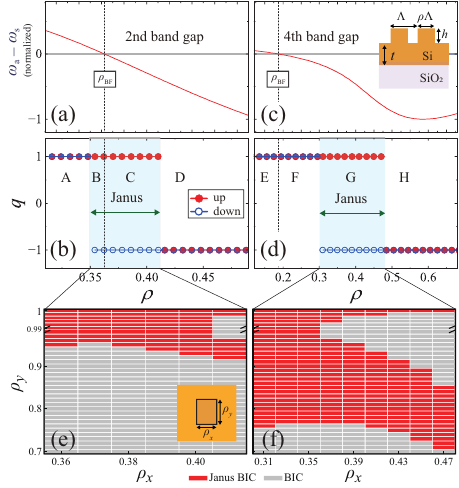}
\caption{Janus BICs in silicon ZCGs on SiO$_2$ with an air cover, calculated by RCWA. (a) Frequency gap $\omega_{\mathrm a}-\omega_{\mathrm s}$ at the second stop band and (b) channel charges for $t=0.1\Lambda$ and $h=0.4\Lambda$. (c),(d) The same at the fourth stop band for $t=0.45\Lambda$ and $h=0.2\Lambda$. (e),(f) Janus regions of 2D ZCGs with rectangular Si pillars of widths $\rho_x\Lambda$ and $\rho_y\Lambda$ at the second and fourth stop bands, respectively, with the same $t$ and $h$ as in (b) and (d). The $\rho_y$ scale is magnified fourfold above the axis break at $0.99$. A--D in (b) and E--H in (d) mark the regions separated by the band flip and the two channel flips.}
\label{fig6}
\end{figure}

\textit{Zero-contrast gratings}---As Eq.~\eqref{eq:width} and Table~\ref{tab1} show, broad band-flip-induced Janus windows require strong radiative coupling and a large difference between the channel phases. A ZCG~\cite{Magnusson2014,ShokoohSaremi2014}, in which ridges stand on a slab of the same material, can provide both conditions. The large index contrast between the Si ridges and the air gaps strengthens the radiative coupling. The ZCG lets the upward and downward channels radiate through the ridges and the slab, respectively, so that $\tan\theta^{\sigma}$ can differ substantially between the two channels. Figures~\ref{fig6}(a) and \ref{fig6}(b) show the gap and the channel charges of a silicon ZCG on SiO$_2$ with an air cover at the second stop band. The gap closes at the band flip, $\rho_{\mathrm{BF}}=0.36258$. The two channel charges, however, flip at different values of $\rho$. The downward channel flips at $\rho=0.35048$, whereas the upward channel flips at $\rho=0.41255$, on opposite sides of the band flip, thereby opening a Janus window between them. At the fourth stop band, as shown in Figs.~\ref{fig6}(c) and \ref{fig6}(d), the band flips at $\rho_{\mathrm{BF}}=0.18730$, but neither channel flips there. The charges flip at $\rho=0.30147$ in the downward channel and at $\rho=0.48066$ in the upward channel, yielding a broad Janus window between the two channel-flip points. Janus BICs can be realized experimentally in silicon ZCGs. Around $\lambda=1.55~\mu$m, they emerge at the second stop band for $\Lambda=0.70~\mu$m and ridge widths of $245$--$289$ nm, and at the fourth stop band for $\Lambda=0.94~\mu$m and ridge widths of $283$--$452$ nm, dimensions readily achievable with standard silicon fabrication techniques.

\begin{figure}[t]
\centering
\includegraphics[width=\columnwidth]{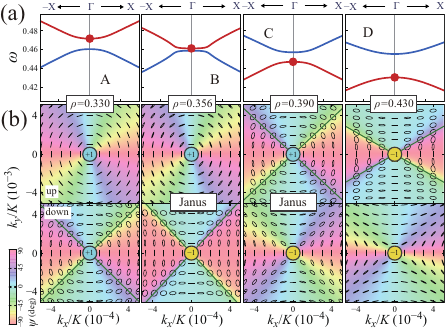}
\caption{Band flip and charge flips in the regions A--D of Fig.~\ref{fig6}(b), calculated by RCWA. (a) Dispersion of the BIC band (red) and the leaky band (blue) along $k_x$ for $|k_x|\le0.08K$, with $\omega$ in units of $2\pi c/\Lambda$. The dot marks the BIC at $\Gamma$. (b) Far-field polarization maps of the upward (top) and downward (bottom) channels. Colors show the polarization orientation angle $\psi$, and green lines mark the directions in which $|E_x^{\sigma}|=|E_y^{\sigma}|$ in the channel near its flip.}
\label{fig7}
\end{figure}

Figures~\ref{fig6}(e) and \ref{fig6}(f) show 2D ZCGs, in which rectangular Si pillars, $\rho_x\Lambda$ wide and $\rho_y\Lambda$ long, form a square lattice on the same slabs. At $\rho_y=1$, the pillars merge into the ridges of the 1D ZCGs. For the BIC that reduces to the 1D BIC at $\rho_y=1$, the lower edge of the Janus region lies at $\rho_y\approx0.92$--$0.96$ at the second stop band and at $\rho_y\approx0.71$--$0.77$ at the fourth stop band, depending on $\rho_x$. Janus BICs therefore persist when the continuous ridges are broken into discrete pillars.

Figure~\ref{fig7} tracks the 1D ZCG at the second stop band through the regions A--D marked in Fig.~\ref{fig6}(b). As Fig.~\ref{fig7}(a) shows, the BIC lies at the upper band edge in A and B and at the lower band edge in C and D, placing the band flip between B and C. Figure~\ref{fig7}(b) shows the far-field polarization of the two channels. Both channels carry $q=+1$ in A. The downward channel flips while the gap is still open, so that the BIC in B is already a Janus BIC at the upper band edge. Across the band flip, the BIC moves to the lower band edge, but both charges remain unchanged in C. The upward channel flips between C and D, after which both channels carry $q=-1$. In each region, the channel closer to its charge flip forms lines of nearly circular polarization along the directions in which $|E_x^{\sigma}|=|E_y^{\sigma}|$, because $\cos\delta^{\sigma}$ passes through zero at the charge flip of channel $\sigma$. At the fourth stop band, the regions E--H in Fig.~\ref{fig6}(d) follow a different order. The band flip occurs between E and F while both channels keep $q=+1$, and the channels flip only afterward, the downward channel between F and G and the upward channel between G and H, so that the Janus BIC in G lies entirely at the lower band edge. Taken together, the results in Figs.~\ref{fig6} and \ref{fig7} demonstrate that, in vertically asymmetric lattices, charge flips are governed jointly by the band gap and the phases of the upward and downward channels.

\textit{Two-dimensional lattices}---In the Supplemental Material~\cite{SM}, we extend the coupled-mode model to a square lattice of rectangular pillars and derive an analytical expression for the topological charge of the symmetry-protected BICs. The band edge at the $\Gamma$ point hosts four modes, $x$- and $y$-polarized BICs and their leaky partners. Near each BIC, the far field again takes the form $(c^{\sigma}k_y,\,d^{\sigma}k_x)$, so that the charge is $q^{\sigma}=-\mathrm{sgn}\,\mathrm{Re}(c^{\sigma*}d^{\sigma})$. Writing $E_x^{\sigma}+iE_y^{\sigma}=\mathcal A^{\sigma}k_++\mathcal B^{\sigma}k_-$, with $k_\pm=k_x\pm ik_y$, the charge becomes
\begin{equation}\label{eq:qAB}
q^{\sigma}=\mathrm{sgn}\big(|\mathcal A^{\sigma}|-|\mathcal B^{\sigma}|\big).
\end{equation}
The charge is therefore $+1$ or $-1$ and flips where $|\mathcal A^{\sigma}|=|\mathcal B^{\sigma}|$. For square pillars ($\rho_x=\rho_y$), fourfold rotational symmetry forces $\mathcal A^{\sigma}$ or $\mathcal B^{\sigma}$ to vanish, depending on the symmetry of the BIC, so that this condition cannot be met and the charge remains $-1$ or $+1$, respectively, in both channels. Rectangular pillars ($\rho_x\neq\rho_y$) lower the symmetry to $C_{2v}$ and allow both coefficients to be nonzero. The charge then flips either where the BIC crosses a leaky mode, as at a band flip, or where off-$\Gamma$ BICs or C points merge with the BIC while the gaps remain open. In a vertically asymmetric lattice, $\mathcal A^{\sigma}$ and $\mathcal B^{\sigma}$ differ between the two channels, so that the upward and downward channels flip at different values of $\rho_x$ and $\rho_y$, producing the Janus regions in Figs.~\ref{fig6}(e) and \ref{fig6}(f).

\end{document}


\title{Supplemental Material for ``Channel-Differentiated Topological Charge Flips and Formation of Janus Bound States in the Continuum in Single-Layer Photonic Lattices''}
\author{Sun-Goo Lee}
\affiliation{Department of Data Information and Physics, Kongju National University, Gongju, 32588, South Korea}
\affiliation{Institute of Application and Fusion for Light, Kongju National University, Cheonan, 31080, South Korea}

\author{Wook-Jae Lee}
\affiliation{Department of Data Information and Physics, Kongju National University, Gongju, 32588, South Korea}
\affiliation{Institute of Application and Fusion for Light, Kongju National University, Cheonan, 31080, South Korea}
\date{\today}
\makeatletter\let\frontmatter@abstract@produce\relax\makeatother
\maketitle
\vspace{\baselineskip}

\section{S1. Coupled-mode model and dispersion relation}
\subsection{Lattice and harmonic equations}
Figure~\ref{fig:band}(a) shows the 1D lattice. The grating layer $|z|<t/2$, with alternating regions $\epsilon_{\mathrm h}$ and $\epsilon_{\mathrm l}$ of period $\Lambda$ and duty cycle $\rho$, is bounded by a cover $\epsilon_{\mathrm c}$ ($z>t/2$) and a substrate $\epsilon_{\mathrm s}$ ($z<-t/2$). As in the main text, the dielectric function is expanded as
\begin{equation}\label{eq:eps}
\epsilon(x,z)=\epsilon_0(z)+\sum_{n\ge1}\epsilon_n(z)\cos(nKx),
\end{equation}
where $\epsilon_0(z)$ equals $\epsilon_{\mathrm c}$, $\epsilon_{\mathrm{avg}}$, and $\epsilon_{\mathrm s}$ in the cover, the grating layer, and the substrate, respectively, and $\epsilon_n=(2\Delta\epsilon/n\pi)\sin(n\pi\rho)$ in the grating layer and zero outside. Moving the modulation to the right-hand side, Eq.~(1) of the main text becomes
\begin{equation}\label{eq:wave}
\nabla\times\nabla\times\mathbf E-k_0^2\epsilon_0(z)\mathbf E=k_0^2\sum_{n\ge1}\epsilon_n(z)\cos(nKx)\,\mathbf E.
\end{equation}
With the Bloch expansion $\mathbf E=\sum_m\mathbf E_m(z)e^{i(k_x+mK)x+ik_yy}$, the harmonic $m$ travels along $\mathbf q_m=(k_x+mK,k_y)$ and, because $\cos(nKx)=(e^{inKx}+e^{-inKx})/2$, is driven by $(k_0^2/2)\mathbf P_m$ with
\begin{equation}\label{eq:Pm}
\mathbf P_m=\sum_{n\ge1}\epsilon_n\big(\mathbf E_{m-n}+\mathbf E_{m+n}\big).
\end{equation}
Each harmonic splits into a TE part along $\ex_m$ and a TM part along $\px_m$,
\begin{equation}
\ex_m=\hat{\mathbf z}\times\px_m,\qquad \px_m=s_m\mathbf q_m/|\mathbf q_m|,
\end{equation}
where $s_m=\sgn m$ orients the two vectors so that $\ex_m\to\hat{\mathbf y}$ at the $\Gamma$ point for every $m\neq0$, and $s_0=1$. Near the $\Gamma$ point, $\ex_m\simeq(-k_y/mK,1)$ and $\px_m\simeq(1,k_y/mK)$, so that $\ex_{\pm2}$ are the polarization vectors $\ex_\pm$ of the main text. Except for the TE part of the guided waves, each harmonic responds to its source as
\begin{equation}\label{eq:resp}
E_{m,e}=-\frac{k_0^2}{2}G_m[P_{m,e}],\qquad E_{m,p}=-\frac{k_0^2}{2}G_m^{\TM}[P_{m,p}],
\end{equation}
where $G_m$ is the Green function of $L_m=d^2/dz^2+\epsilon_0(z)k_0^2-|\mathbf q_m|^2$ with TE interface conditions, and $G_m^{\TM}$ solves $L_mu=(\kappa^2/\epsilon_0k_0^2)f$ with $u$ and $(\epsilon_0/\kappa^2)\,du/dz$ continuous, $\kappa^2=\epsilon_0k_0^2-|\mathbf q_m|^2$ in each region. Near the Bragg frequency, $|\mathbf q_m|$ with $m\neq0$ exceeds $\sqrt{\epsilon_{\mathrm c}}k_0$ and $\sqrt{\epsilon_{\mathrm s}}k_0$, so that $G_m$ and $G_m^{\TM}$ give fields bound to the grating layer. The Green function $G_0$ of the zeroth-order harmonic is outgoing, radiating plane waves into the cover and the substrate.

\subsection{Field approximation}
Near the third-order $\Gamma$ point, the fourth stop band is formed by the two counterpropagating TE$_0$ guided waves of the $\pm2K$ harmonics. Retaining Fourier components through $\epsilon_4$, we approximate the electric field as in Eq.~(2) of the main text,
\begin{multline}\label{eq:field}
\mathbf E=\Big[\big(A\hat{\mathbf e}_+e^{i2Kx}+B\hat{\mathbf e}_-e^{-i2Kx}\big)\phi+\mathbf E_{\mathrm{rad}}\\
+\sum_{m=1,3}\big(\mathbf E_{+m}e^{imKx}+\mathbf E_{-m}e^{-imKx}\big)\Big]e^{i(k_xx+k_yy)}.
\end{multline}
Here, $A$ and $B$ are the amplitudes of the counterpropagating guided waves forming the $\pm2K$ harmonics, $\phi$ is the transverse profile of the TE$_0$ mode of the unmodulated reference slab $\epsilon_0(z)$, and $\mathbf E_{\mathrm{rad}}=\mathbf E_0$ and $\mathbf E_{\pm m}$ are the $z$-dependent amplitudes of the zeroth-order and intermediate harmonics. The guided waves are of zeroth order in the modulation, and the other harmonics are their responses, Eq.~\eqref{eq:resp}, of first order. Below, $\mathbf E_{\pm2}$ denotes the $\pm2K$ harmonic and $(A_+,A_-)=(A,B)$.

\emph{Intermediate harmonics.} By Eq.~\eqref{eq:Pm}, the $\pm K$ and $\pm3K$ harmonics are driven by
\begin{equation}\label{eq:P13}
\mathbf P_{\pm1}=\epsilon_1\mathbf E_{\pm2}+\epsilon_3\mathbf E_{\mp2}+\epsilon_1\mathbf E_{\mathrm{rad}},\qquad
\mathbf P_{\pm3}=\epsilon_1\mathbf E_{\pm2}+\epsilon_3\mathbf E_{\mathrm{rad}}.
\end{equation}
Couplings among the intermediate harmonics, such as $\epsilon_2\mathbf E_{\mp1}$, are omitted, because they change the coupling coefficients only at higher order in the modulation. The guided-wave terms are of first order in the modulation. Projecting them on $\ex_m$ and $\px_m$, Eq.~\eqref{eq:resp} gives
\begin{align}\label{eq:Em}
\mathbf E_m=-\frac{k_0^2}{2}\Big\{&\epsilon_{|m-2|}A\Big[\ex_m(\ex_m\cdot\ex_+)G_m+\px_m(\px_m\cdot\ex_+)G_m^{\TM}\Big][\phi]\nonumber\\
+\,&\epsilon_{|m+2|}B\Big[\ex_m(\ex_m\cdot\ex_-)G_m+\px_m(\px_m\cdot\ex_-)G_m^{\TM}\Big][\phi]\Big\},\qquad m=\pm1,\pm3,
\end{align}
where the $\pm3K$ harmonics are driven only by the guided wave on their own side, through $\epsilon_1$. The TM parts appear because $\ex_\pm$ is not perpendicular to $\mathbf q_m$ for $k_y\neq0$, and they are of first order in $k_y$. The $\mathbf E_{\mathrm{rad}}$ terms of Eq.~\eqref{eq:P13} are of second order in the modulation and return the radiation to the guided waves through the intermediate harmonics.

\emph{TM part of the $\pm2K$ harmonics.} The $\pm2K$ harmonic is also driven by $\epsilon_4\mathbf E_{\mp2}$. Its TE part couples the two guided waves. Its TM part, of first order in $k_y$, gives the bound TM field
\begin{equation}\label{eq:E2p}
\mathbf E_{\pm2,p}=-\frac{k_0^2}{2}\epsilon_4A_\mp(\px_{\pm2}\cdot\ex_\mp)\,\px_{\pm2}\,G_{\pm2}^{\TM}[\phi],
\end{equation}
so that the $\pm2K$ harmonic consists of the guided wave and its TM response, $\mathbf E_{\pm2}=A_\pm\ex_\pm\phi+\mathbf E_{\pm2,p}$.

\emph{Radiating harmonic.} The zeroth-order harmonic is driven by
\begin{equation}\label{eq:P0}
\mathbf P_0=\epsilon_2(\mathbf E_2+\mathbf E_{-2})+\epsilon_1(\mathbf E_1+\mathbf E_{-1})+\epsilon_3(\mathbf E_3+\mathbf E_{-3}),
\end{equation}
so that $\mathbf E_{\mathrm{rad}}=-(k_0^2/2)G_0[\mathbf P_0]$, and $\mathbf P_0$ is the source $\mathbf S$ of the main text.

\emph{Back action on the guided waves.} The $\pm2K$ harmonics are driven by
\begin{equation}\label{eq:P2}
\mathbf P_{\pm2}=\epsilon_4\mathbf E_{\mp2}+\epsilon_2\mathbf E_{\mathrm{rad}}+\epsilon_1\big(\mathbf E_{\pm1}+\mathbf E_{\pm3}\big)+\epsilon_3\mathbf E_{\mp1}.
\end{equation}
The $\epsilon_4$ term couples the guided waves at first order in the modulation. The other terms are of second order. They return to the guided waves the bound responses of the intermediate harmonics and, together with the $\mathbf E_{\mathrm{rad}}$ terms of Eq.~\eqref{eq:P13}, the radiation.

\subsection{Projection on the guided waves}
We use the normalized frequency $\Omega=k_0/K$. At the Bragg frequency $\Omega_0$ of the reference slab, the TE$_0$ mode has the propagation constant $2K$, and its profile satisfies $[d^2/dz^2+\epsilon_0(z)K^2\Omega_0^2-4K^2]\phi=0$. Inserting the guided wave $A\ex_+\phi$ into the TE part of the $m=+2$ equation, multiplying by $\phi$, and integrating over $z$ gives, to first order in $\Omega-\Omega_0$,
\begin{equation}\label{eq:proj}
A\Big[2K^2\Omega_0(\Omega-\Omega_0)\!\int\!\epsilon_0\phi^2dz-\big(|\mathbf q_{+2}|^2-4K^2\big)\!\int\!\phi^2dz\Big]=-\frac{k_0^2}{2}\int\phi\,P_{+2,e}\,dz,
\end{equation}
where we used $k_0^2-K^2\Omega_0^2\simeq2K^2\Omega_0(\Omega-\Omega_0)$. Dividing by $-N_\phi=-4K\int\phi^2dz$ and treating the $m=-2$ equation in the same way gives, for any in-plane $\mathbf k$,
\begin{equation}\label{eq:KHgen}
Kn_{\mathrm g}(\Omega_0-\Omega)\begin{pmatrix}A\\B\end{pmatrix}=\begin{pmatrix}h_{AA}-\Delta_+&h_{AB}\\h_{BA}&h_{BB}-\Delta_-\end{pmatrix}\begin{pmatrix}A\\B\end{pmatrix},
\end{equation}
where $\Delta_\pm=(|\mathbf q_{\pm2}|^2-4K^2)/4K=\pm k_x+(k_x^2+k_y^2)/4K$ is the exact detuning of each guided wave. Its part $(k_x^2+k_y^2)/4K$ is common to the two guided waves and shifts both band edges equally, and dropping it gives Eq.~(3) of the main text. The diagonal coefficients $h_{AA}(\mathbf k)$ and $h_{BB}(\mathbf k)$ and the off-diagonal coefficients $h_{AB}(\mathbf k)$ and $h_{BA}(\mathbf k)$ are $(k_0^2/2N_\phi)\int\phi\,\ex_\pm\cdot\mathbf P_{\pm2}\,dz$ evaluated with the source produced by the same and by the opposite guided wave of unit amplitude, respectively, through all paths of Eqs.~\eqref{eq:P13}--\eqref{eq:P2}, including the radiation through $\mathbf E_{\mathrm{rad}}$. By the symmetry of the Green functions (reciprocity), the coupling matrix is symmetric, $h_{AB}=h_{BA}$. The mirror symmetry $x\to-x$ interchanges the two guided waves and reverses $k_x$, so that $h_{AA}(k_x,k_y)=h_{BB}(-k_x,k_y)$. Hence $h_{AA}=h_{BB}$ on the $k_y$ axis. Along $k_x$, their difference, which is odd in $k_x$, changes the detuning $\pm k_x$ only at higher order in the modulation, as noted below Eq.~\eqref{eq:hr}, and is neglected in the main text. At the $\Gamma$ point, every harmonic is $y$ polarized, the TM parts vanish, and $G_{-m}=G_m$. Equations~\eqref{eq:Em} and \eqref{eq:P0} then give the source of the radiating harmonic,
\begin{equation}\label{eq:St}
\mathbf P_0=(A+B)\tilde S_y\hat{\mathbf y},\qquad
\tilde S_y=\epsilon_2\phi-\frac{k_0^2}{2}\epsilon_1\big[(\epsilon_1+\epsilon_3)G_1+\epsilon_3G_3\big][\phi],
\end{equation}
and $\mathbf E_{\mathrm{rad}}=-(k_0^2/2)(A+B)G_0[\tilde S_y]\hat{\mathbf y}$. Inserting Eqs.~\eqref{eq:P13}, \eqref{eq:Em}, and \eqref{eq:P0} into Eq.~\eqref{eq:P2}, and using the symmetry of $G_1$ and $G_3$, the $\mathbf E_{\mathrm{rad}}$ terms collect into $\int\tilde S_y E_{\mathrm{rad},y}\,dz$, so that the radiation leaves and returns to the guided waves through the same source. At the $\Gamma$ point, $h_{AA}=h_{BB}=ih_{\mathrm r}$ and $h_{AB}=h_{\mathrm g}+ih_{\mathrm r}$, and Eq.~(3) of the main text therefore becomes
\begin{equation}\label{eq:KH}
Kn_{\mathrm g}(\Omega_0-\Omega)\begin{pmatrix}A\\B\end{pmatrix}=
\begin{pmatrix}ih_{\mathrm r}-k_x& h_{\mathrm g}+ih_{\mathrm r}\\ h_{\mathrm g}+ih_{\mathrm r}& ih_{\mathrm r}+k_x\end{pmatrix}\begin{pmatrix}A\\B\end{pmatrix},
\end{equation}
with
\begin{align}
n_{\mathrm g}&=\frac{\Omega_0\int\epsilon_0\phi^2dz}{2\int\phi^2dz},\qquad N_\phi=4K\!\int\!\phi^2dz,\label{eq:h0}\\
h_{\mathrm g}&=\frac{k_0^2}{2N_\phi}\Big(\epsilon_4\!\int\!\phi^2dz+\frac{k_0^2}{2}\int\phi\,\big[(\epsilon_1-\epsilon_3)^2G_1+\epsilon_1^2G_3\big][\phi]\,dz\Big),\label{eq:hg}\\
h_{\mathrm r}&=\frac{ik_0^4}{4N_\phi}\Big(\int\tilde S_y\,G_0[\tilde S_y]\,dz+\int\phi\,\big[(\epsilon_1^2+\epsilon_3^2)G_1+\epsilon_1^2G_3\big][\phi]\,dz\Big),\label{eq:hr}
\end{align}
where $n_{\mathrm g}$ is the group index of the guided wave, so that $d\Omega/dk_x=1/(Kn_{\mathrm g})$ in the absence of coupling. The integrals in $n_{\mathrm g}$ and $N_\phi$ run over all $z$, and those containing $\epsilon_n$ over the grating layer. The first term of $h_{\mathrm r}$ comes from the radiation through $\mathbf E_{\mathrm{rad}}$, the $\epsilon_4$ term of $h_{\mathrm g}$ from the first-order coupling of $+2K$ and $-2K$, and the $G_1$ and $G_3$ terms from the bound paths through the $\pm K$ and $\pm3K$ harmonics. The coefficients do not depend on the normalization of $\phi$, and their dependence on $\mathbf k$ changes the $k_x$ terms of Eq.~\eqref{eq:KH} only at higher order in the modulation.

\subsection{Dispersion relation}
The eigenvalues of Eq.~\eqref{eq:KH} give the dispersion relation, Eq.~(4) of the main text,
\begin{equation}\label{eq:disp}
\Omega(k_x)=\Omega_0-\frac{ih_{\mathrm r}\pm\sqrt{k_x^2+(h_{\mathrm g}+ih_{\mathrm r})^2}}{Kn_{\mathrm g}}.
\end{equation}
At $k_x=0$, the antisymmetric state $A=-B$, for which $\mathbf P_0=0$ by Eq.~\eqref{eq:St} and the radiation vanishes, has the real frequency $\Omega_{\mathrm a}=\Omega_0+h_{\mathrm g}/(Kn_{\mathrm g})$ and is the symmetry-protected bound state in the continuum (BIC). The symmetric state $A=B$ is the leaky state at $\Omega_{\mathrm s}=\Omega_0-(h_{\mathrm g}+2ih_{\mathrm r})/(Kn_{\mathrm g})$, whose radiative loss $-\Ima\Omega_{\mathrm s}=2\Rea h_{\mathrm r}/(Kn_{\mathrm g})$ is set by $\Rea h_{\mathrm r}$ and whose real part is shifted by $\Ima h_{\mathrm r}$. The band gap $|\Rea(\Omega_{\mathrm a}-\Omega_{\mathrm s})|=2|h_{\mathrm g}-\Ima h_{\mathrm r}|/(Kn_{\mathrm g})$ closes at $G_{\mathrm{band}}\equiv h_{\mathrm g}-\Ima h_{\mathrm r}=0$, across which the BIC moves to the opposite band edge in a band flip. At the band flip, $h_{\mathrm g}+ih_{\mathrm r}=i\Rea h_{\mathrm r}$, and Eq.~\eqref{eq:disp} becomes
\begin{equation}\label{eq:dispBF}
\Omega(k_x)=\Omega_0-\frac{ih_{\mathrm r}\pm\sqrt{k_x^2-(\Rea h_{\mathrm r})^2}}{Kn_{\mathrm g}},
\end{equation}
so that the two eigenfrequencies and their eigenvectors coalesce at the exceptional points $k_x=\pm k_{\mathrm{EP}}$ with $k_{\mathrm{EP}}=|\Rea h_{\mathrm r}|$.

\begin{figure}[t]
\includegraphics{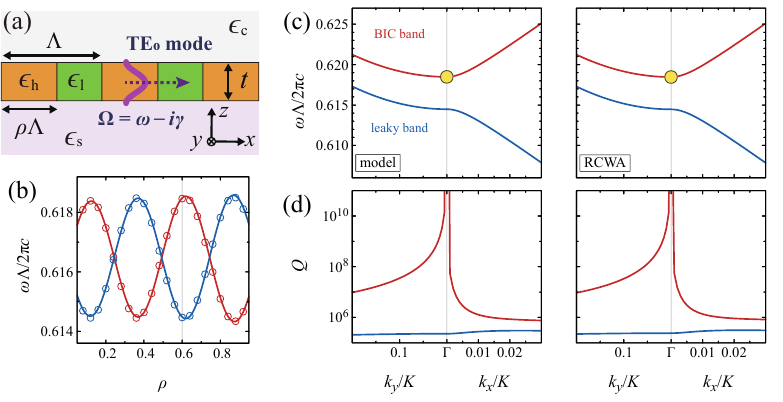}
\caption{\label{fig:band}(a) Schematic of the 1D lattice. (b)--(d) Results for the vertically symmetric lattice ($\epsilon_{\mathrm c}=\epsilon_{\mathrm s}=2.1316$, $\epsilon_{\mathrm{avg}}=12$, $\Delta\epsilon=1$, $t=0.5\Lambda$), calculated from the model and by RCWA. (b) Frequencies of the BIC (red) and the leaky state (blue) at the $\Gamma$ point, with the model as lines and RCWA as circles. (c) Bands and (d) $Q$ factors near the $\Gamma$ point at $\rho=0.6$, marked by the gray line in (b).}
\end{figure}

Figures~\ref{fig:band}(b)--\ref{fig:band}(d) compare the model with rigorous coupled-wave analysis (RCWA) for the vertically symmetric lattice of Fig.~2 of the main text ($\epsilon_{\mathrm c}=\epsilon_{\mathrm s}=2.1316$, $\epsilon_{\mathrm{avg}}=12$, $\Delta\epsilon=1$, $t=0.5\Lambda$). The model results are calculated as described in Sec.~S6, with the frequency iterated to the eigenfrequency of each band. Figure~\ref{fig:band}(b) shows that the BIC at the $\Gamma$ point moves between the upper and lower band edges at the three band flips $\rho_{\mathrm{F1}}$, $\rho_{\mathrm{F2}}$, and $\rho_{\mathrm{F5}}$ of the main text, and the two calculations agree well. Figures~\ref{fig:band}(c) and \ref{fig:band}(d) show the bands and the $Q$ factors near the $\Gamma$ point at $\rho=0.6$, which also agree well between the two calculations.

\subsection{BIC branch near the $\Gamma$ point}
For $k_x\neq0$, the detuning $\pm k_x$ makes the amplitudes of the two guided waves unequal. Treating the detuning as a perturbation of Eq.~\eqref{eq:KH}, with the eigenvalues $-h_{\mathrm g}$ and $h_{\mathrm g}+2ih_{\mathrm r}$ of the antisymmetric and symmetric states, gives to first order in $k_x$
\begin{equation}\label{eq:ABk}
\begin{pmatrix}A\\B\end{pmatrix}=a\begin{pmatrix}1\\-1\end{pmatrix}+\frac{ak_x}{2(h_{\mathrm g}+ih_{\mathrm r})}\begin{pmatrix}1\\1\end{pmatrix},
\end{equation}
where $A=-B=a$ at the $\Gamma$ point, so that $A+B=ak_x/(h_{\mathrm g}+ih_{\mathrm r})$ and $A-B=2a$, as used in the main text. At the band flip, $A+B=-iak_x/\Rea h_{\mathrm r}$.

\section{S2. Radiated field and topological charge}
\subsection{Radiation into the cover and the substrate}
To first order in $\mathbf k$, the radiated field follows from the zeroth-order equation at the $\Gamma$ point, $\mathbf E_{\mathrm{rad}}=-(k_0^2/2)G_0[\mathbf S]$, for both polarizations, where $G_0$ is the outgoing Green function of $d^2/dz^2+\epsilon_0(z)k_0^2$. In the grating layer, the solutions that are outgoing in the cover and in the substrate are
\begin{equation}\label{eq:V}
V_{\mathrm{up}}=\cos\mu\big(z-\tfrac t2\big)+\frac{i\mu_{\mathrm c}}{\mu}\sin\mu\big(z-\tfrac t2\big),\qquad
V_{\mathrm{dn}}=\cos\mu\big(z+\tfrac t2\big)-\frac{i\mu_{\mathrm s}}{\mu}\sin\mu\big(z+\tfrac t2\big),
\end{equation}
with $\mu=\sqrt{\epsilon_{\mathrm{avg}}}\,k_0$, $\mu_{\mathrm c}=\sqrt{\epsilon_{\mathrm c}}\,k_0$, and $\mu_{\mathrm s}=\sqrt{\epsilon_{\mathrm s}}\,k_0$, so that $G_0(z,z')=V_{\mathrm{up}}(z_>)V_{\mathrm{dn}}(z_<)/W$ with the Wronskian $W=V_{\mathrm{dn}}V_{\mathrm{up}}'-V_{\mathrm{dn}}'V_{\mathrm{up}}$. The field in the cover is thus an outgoing plane wave with amplitude $-(k_0^2/2)\int V_{\mathrm{dn}}\mathbf S\,dz/W$ at $z=t/2$, and the field in the substrate an outgoing plane wave with amplitude $-(k_0^2/2)\int V_{\mathrm{up}}\mathbf S\,dz/W$ at $z=-t/2$. Writing $V_{\mathrm{dn}}$ and $V_{\mathrm{up}}$ as waves traveling in the layer, the field radiated into channel $\sigma$ takes the form used in the main text,
\begin{equation}\label{eq:Esig}
\mathbf E_{\mathrm{rad}}^\sigma=-\frac{k_0^2}{2}\mathcal K^\sigma\!\int u^\sigma\mathbf S\,dz,
\end{equation}
with the radiation profiles and coupling constants
\begin{equation}\label{eq:uK}
u^\uparrow=e^{-i\mu z}+r_{\mathrm s}e^{i\mu t}e^{i\mu z},\qquad
u^\downarrow=e^{i\mu z}+r_{\mathrm c}e^{i\mu t}e^{-i\mu z},\qquad
\mathcal K^\sigma=\frac{\tau^\sigma e^{i\mu t/2}}{2i\mu D}.
\end{equation}
Here, $r_{\mathrm c,\mathrm s}=(\mu-\mu_{\mathrm c,\mathrm s})/(\mu+\mu_{\mathrm c,\mathrm s})$ are the reflection coefficients at the grating--cover and grating--substrate interfaces, $\tau^\uparrow=2\mu/(\mu+\mu_{\mathrm c})$ and $\tau^\downarrow=2\mu/(\mu+\mu_{\mathrm s})$ are the transmission coefficients into the cover and the substrate, and $D=1-r_{\mathrm c}r_{\mathrm s}e^{2i\mu t}$ accounts for the multiple reflections in the layer. The profile $u^\uparrow$ collects the wave emitted upward and the wave emitted downward and reflected at the substrate, and $u^\downarrow$ collects the reverse, so that $u^\uparrow\propto V_{\mathrm{dn}}$ and $u^\downarrow\propto V_{\mathrm{up}}$. At $\mathbf k\neq0$, the TE part ($\ex_0$) and the TM part ($\px_0$) of $\mathbf S$ radiate with their own profiles and coupling constants, obtained from Eq.~\eqref{eq:uK} with $\mu$, $\mu_{\mathrm c}$, and $\mu_{\mathrm s}$ evaluated at the in-plane wave number $|\mathbf k|$ and, for TM, with the TM interface conditions. Although $\ex_0$ and $\px_0$ are not defined at $\mathbf k=0$, the TE and TM interface conditions coincide at normal incidence, so that the field radiated at $\mathbf k\neq0$ tends continuously to Eq.~\eqref{eq:Esig} as $\mathbf k\to0$, and the charge of the model is evaluated on loops with $\mathbf k\neq0$.

\subsection{Radiation source}
At the $\Gamma$ point, the BIC has $A=-B$, and Eq.~\eqref{eq:St} gives $\mathbf P_0=0$, so that the BIC does not radiate. To obtain the $\mathbf k\to0$ limits of the radiation factors of Eq.~(8) of the main text, we evaluate the source $\mathbf S=\mathbf P_0$ of Eq.~\eqref{eq:P0} to first order in $\mathbf k$, with the Green functions evaluated at the $\Gamma$ point, $|\mathbf q_m|=|m|K$. The $\mathbf k$ dependence of the Green functions and of the profiles adds terms of higher order in the modulation, which are kept on the two symmetry axes at the end of this subsection. To first order in $k_y$, the polarization vectors of Sec.~S1 give $\ex_m\cdot\ex_\pm\simeq1$ and
\begin{equation}\label{eq:pe}
\hat e_{m,x}\simeq-\frac{k_y}{mK},\qquad \hat p_{m,x}\simeq1,\qquad \px_m\cdot\ex_\pm\simeq\frac{k_y}{K}\Big(\frac1m\mp\frac12\Big).
\end{equation}
The $y$ component of $\mathbf S$ is then carried by the TE parts alone and is proportional to $A+B$, as in Eq.~\eqref{eq:St}, because the TM parts enter it only at second order in $k_y$. The $x$ component arises from the $x$ components of $\ex_\pm$ and $\ex_m$ and from the TM parts, Eqs.~\eqref{eq:Em} and \eqref{eq:E2p}. Inserting Eq.~\eqref{eq:pe} into Eqs.~\eqref{eq:Em}, \eqref{eq:E2p}, and \eqref{eq:P0}, the three terms of $P_{0,x}$ become
\begin{align}\label{eq:P0x}
\epsilon_2\big(E_{2,x}+E_{-2,x}\big)&=-(A-B)\frac{k_y}{2K}\,\epsilon_2\big(\phi-k_0^2\epsilon_4G_2^{\TM}[\phi]\big),\nonumber\\
\epsilon_1\big(E_{1,x}+E_{-1,x}\big)&=(A-B)\frac{k_y}{2K}\,\frac{k_0^2}{2}\epsilon_1\big[\epsilon_1\big(2G_1-G_1^{\TM}\big)+\epsilon_3\big(3G_1^{\TM}-2G_1\big)\big][\phi],\\
\epsilon_3\big(E_{3,x}+E_{-3,x}\big)&=(A-B)\frac{k_y}{2K}\,\frac{k_0^2}{2}\epsilon_1\epsilon_3\,\tfrac13\big(2G_3+G_3^{\TM}\big)[\phi].\nonumber
\end{align}
The source near the $\Gamma$ point is therefore
\begin{equation}\label{eq:S}
\mathbf S=(A+B)\tilde S_y\hat{\mathbf y}-(A-B)\frac{k_y}{2K}\tilde S_x\hat{\mathbf x},
\end{equation}
with $\tilde S_y$ of Eq.~\eqref{eq:St} and
\begin{equation}\label{eq:Sx}
\tilde S_x=\epsilon_2\big(\phi-k_0^2\epsilon_4G_2^{\TM}[\phi]\big)-\frac{k_0^2}{2}\epsilon_1\Big[\epsilon_1\big(2G_1-G_1^{\TM}\big)+\epsilon_3\Big(3G_1^{\TM}-2G_1+\tfrac13\big(2G_3+G_3^{\TM}\big)\Big)\Big][\phi].
\end{equation}
Because $\phi$ and the bound responses $G_m[\phi]$ and $G_m^{\TM}[\phi]$ are real, $\tilde S_x$ and $\tilde S_y$ are real.

On the two symmetry axes, the source is obtained without expanding the Green functions and the profiles in $\mathbf k$. On the $k_x$ axis ($k_y=0$), every harmonic is $y$ polarized. With the profile of each guided wave taken at its propagation constant $|\mathbf q_{\pm2}|=|2K\pm k_x|$ and denoted by $\phi_\pm$, the $y$ components $S_{\pm,y}$ of the sources of the main text become
\begin{equation}\label{eq:Spmk}
\tilde S_\pm=\epsilon_2\phi_\pm-\frac{k_0^2}{2}\epsilon_1\big[\epsilon_1G_{\pm1}+\epsilon_3\big(G_{\mp1}+G_{\pm3}\big)\big][\phi_\pm],
\end{equation}
where $G_{\pm m}$ is evaluated at $|\mathbf q_{\pm m}|=|mK\pm k_x|$, so that $\mathbf S=(A\tilde S_++B\tilde S_-)\hat{\mathbf y}$, and $\tilde S_\pm$ reduce to $\tilde S_y$ at $k_x=0$. On the $k_y$ axis ($k_x=0$), the mirror symmetry $x\to-x$ gives $|\mathbf q_{-m}|=|\mathbf q_m|=(m^2K^2+k_y^2)^{1/2}$ and keeps $A=-B=a$ at any $k_y$, so that the $y$ components of the two sources cancel and the source is $x$ polarized. Repeating Eq.~\eqref{eq:P0x} with the exact polarization vectors at $k_x=0$ gives $\mathbf S=-(A-B)(k_y/|\mathbf q_2|)\tilde S_x(k_y)\hat{\mathbf x}$ with
\begin{align}\label{eq:Sxky}
\tilde S_x(k_y)={}&\epsilon_2\Big(\phi-\frac{4K^2}{|\mathbf q_2|^2}k_0^2\epsilon_4G_2^{\TM}[\phi]\Big)
-\frac{k_0^2}{2}\epsilon_1\bigg[\epsilon_1\frac{(2K^2+k_y^2)G_1-K^2G_1^{\TM}}{|\mathbf q_1|^2}\nonumber\\
&+\epsilon_3\Big(\frac{3K^2G_1^{\TM}-(2K^2-k_y^2)G_1}{|\mathbf q_1|^2}+\frac{(6K^2+k_y^2)G_3+3K^2G_3^{\TM}}{|\mathbf q_3|^2}\Big)\bigg][\phi],
\end{align}
where $\phi$ is taken at $\beta=|\mathbf q_2|$ and $G_m$ and $G_m^{\TM}$ at $|\mathbf q_m|$. Equation~\eqref{eq:Sxky} reduces to Eq.~\eqref{eq:Sx} at $k_y=0$.

\subsection{Far-field coefficients and radiation factors}
Inserting the sources on the two axes into Eq.~\eqref{eq:Esig}, with $u^\sigma$ and $\mathcal K^\sigma$ taken at the in-plane wave number $k_x$ or $k_y$, gives the $y$-polarized field on the $k_x$ axis and the $x$-polarized field on the $k_y$ axis,
\begin{equation}\label{eq:Fyk}
E_y^\sigma=-\frac{k_0^2}{2}\mathcal K^\sigma(A+B)F_y^\sigma(k_x),\qquad
F_y^\sigma(k_x)=\frac{A\int u^\sigma\tilde S_+dz+B\int u^\sigma\tilde S_-dz}{A+B},
\end{equation}
\begin{equation}\label{eq:Exky}
E_x^\sigma=\frac{k_0^2k_y}{2K}\mathcal K^\sigma aF_x^\sigma(k_y),\qquad F_x^\sigma(k_y)=\frac{2K}{|\mathbf q_2|}\int u^\sigma\tilde S_x(k_y)\,dz.
\end{equation}
With $A-B=2a$ and $A+B=ak_x/(h_{\mathrm g}+ih_{\mathrm r})$ from Eq.~\eqref{eq:ABk}, the limit $k_x\to0$ of $F_y^\sigma(k_x)$ is
\begin{equation}\label{eq:Fy0}
F_y^\sigma(k_x\to0)=\int u^\sigma\tilde S_y\,dz+(h_{\mathrm g}+ih_{\mathrm r})\,\partial_{k_x}\!\int u^\sigma\big(\tilde S_+-\tilde S_-\big)dz\,\Big|_{k_x=0},
\end{equation}
where the second term is smaller than the first by one order in the modulation, and $F_x^\sigma(k_y\to0)=\int u^\sigma\tilde S_x\,dz$. Near the $\Gamma$ point, inserting Eq.~\eqref{eq:S} into Eq.~\eqref{eq:Esig} gives
\begin{equation}\label{eq:EF}
\mathbf E_{\mathrm{rad}}^\sigma=-\frac{k_0^2}{2}\mathcal K^\sigma\Big[(A+B)F_y^\sigma\hat{\mathbf y}-(A-B)\frac{k_y}{2K}F_x^\sigma\hat{\mathbf x}\Big],\qquad
F_j^\sigma=\int u^\sigma\tilde S_j\,dz,
\end{equation}
which, with the second term of Eq.~\eqref{eq:Fy0} added to $F_y^\sigma$, gives the radiation factors of Eq.~(8) of the main text in the limit $\mathbf k\to0$. With $A+B=ak_x/(h_{\mathrm g}+ih_{\mathrm r})$ and $A-B=2a$ from Eq.~\eqref{eq:ABk}, the far field is linear in $\mathbf k$, $\mathbf E_{\mathrm{rad}}^\sigma=(c^\sigma k_y,\,d^\sigma k_x)$, with
\begin{equation}\label{eq:cd}
c^\sigma=\frac{k_0^2}{2K}\,\mathcal K^\sigma a\,F_x^\sigma,\qquad d^\sigma=-\frac{k_0^2}{2}\,\frac{\mathcal K^\sigma a}{h_{\mathrm g}+ih_{\mathrm r}}\,F_y^\sigma,
\end{equation}
which is Eq.~(7) of the main text. The component $k_x$ generates only the $y$-polarized field through $A+B$, and $k_y$ generates only the $x$-polarized field through $A-B$. The two channels share the band coefficients $h_{\mathrm g}$ and $h_{\mathrm r}$ and differ only through $u^\sigma$ and $\mathcal K^\sigma$.

\subsection{Winding number}
On a loop $k_x=\kappa\cos\alpha$, $k_y=\kappa\sin\alpha$ around the $\Gamma$ point, $\mathbf E_{\mathrm{rad}}=(c\kappa\sin\alpha,\,d\kappa\cos\alpha)$, and the Stokes parameters of Eq.~(6) of the main text are
\begin{equation}\label{eq:Stokes}
S_1=\kappa^2\Big[\frac{|c|^2-|d|^2}{2}-\frac{|c|^2+|d|^2}{2}\cos2\alpha\Big],\qquad S_2=\kappa^2\Rea(c^*d)\sin2\alpha.
\end{equation}
For $c,d\neq0$ and $\Rea(c^*d)\neq0$, the point $(S_1,S_2)$ runs twice per loop around an ellipse that encloses the origin, because the semi-axis along $S_1$, $\kappa^2(|c|^2+|d|^2)/2$, exceeds the distance $\kappa^2\big||c|^2-|d|^2\big|/2$ of the center from the origin. Starting on the negative $S_1$ axis at $\alpha=0$, the point turns clockwise for $\Rea(c^*d)>0$ and counterclockwise for $\Rea(c^*d)<0$. The angle $\arg(S_1+iS_2)$ therefore changes by $\mp4\pi$ and $\psi$ by $\mp2\pi$ per loop, and Eq.~(6) of the main text gives
\begin{equation}\label{eq:qcd}
q=-\sgn\Rea(c^*d)=-\sgn\cos\delta,
\end{equation}
where $\delta=\arg d-\arg c$ is the phase difference between $d$ and $c$, which equals the phase difference between $E_y$ and $E_x$ in the first quadrant of the far field. The winding does not depend on the loop radius or on a linear rescaling of the axes. At $\Rea(c^*d)=0$ with $c,d\neq0$, the ellipse collapses to a segment through the origin, the far field is circularly polarized along the two directions $|ck_y|=|dk_x|$, and the charge changes sign. At $c=0$ or $d=0$, the linear far field vanishes along a whole axis, as at the merging points of Sec.~S3.

\subsection{Charge of each channel}
Inserting Eq.~\eqref{eq:cd} into Eq.~\eqref{eq:qcd}, with
\begin{equation}
c^{\sigma*}d^\sigma=-\frac{k_0^4}{4K}\,\big|\mathcal K^\sigma a\big|^2\,\frac{F_x^{\sigma*}F_y^\sigma}{h_{\mathrm g}+ih_{\mathrm r}},
\end{equation}
gives the charge of channel $\sigma$,
\begin{equation}\label{eq:qF}
q^\sigma=\sgn\Rea\bigg[\frac{F_x^{\sigma*}F_y^\sigma}{h_{\mathrm g}+ih_{\mathrm r}}\bigg].
\end{equation}
Because $\tilde S_x$ and $\tilde S_y$ are real, the phases of $F_x^\sigma$ and $F_y^\sigma$ come, to leading order in the modulation, only from the radiation profile $u^\sigma$. With $h_{\mathrm g}+ih_{\mathrm r}=G_{\mathrm{band}}+i\Rea h_{\mathrm r}$,
\begin{equation}
\Rea\bigg[\frac{F_x^{\sigma*}F_y^\sigma}{h_{\mathrm g}+ih_{\mathrm r}}\bigg]=\frac{G_{\mathrm{band}}\Rea\big(F_x^{\sigma*}F_y^\sigma\big)+\Rea h_{\mathrm r}\,\Ima\big(F_x^{\sigma*}F_y^\sigma\big)}{G_{\mathrm{band}}^2+(\Rea h_{\mathrm r})^2},
\end{equation}
which gives Eq.~(9) of the main text,
\begin{equation}\label{eq:q7}
q^\sigma=\sgn\big[G_{\mathrm{band}}\Rea\big(F_x^{\sigma*}F_y^\sigma\big)+\Rea h_{\mathrm r}\,\Ima\big(F_x^{\sigma*}F_y^\sigma\big)\big].
\end{equation}
With $\theta^\sigma=\arg\big(F_x^{\sigma*}F_y^\sigma\big)$, the phase difference $\delta^\sigma$ between $E_y^\sigma$ and $E_x^\sigma$ in the first quadrant satisfies
\begin{equation}\label{eq:cosd}
\cos\delta^\sigma=-\frac{G_{\mathrm{band}}\cos\theta^\sigma+\Rea h_{\mathrm r}\sin\theta^\sigma}{\big[G_{\mathrm{band}}^2+(\Rea h_{\mathrm r})^2\big]^{1/2}},
\end{equation}
whose sign gives the charge, $q^\sigma=-\sgn\cos\delta^\sigma$, and which is plotted in Fig.~4(b) of the main text.

\begin{figure}[t]
\includegraphics{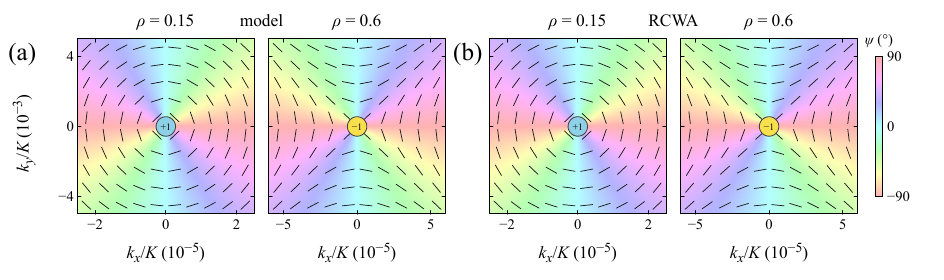}
\caption{\label{fig:vortex}Far-field polarization around the $\Gamma$-point BIC of the vertically symmetric lattice of Fig.~2 of the main text ($\epsilon_{\mathrm c}=\epsilon_{\mathrm s}=2.1316$, $\epsilon_{\mathrm{avg}}=12$, $\Delta\epsilon=1$, $t=0.5\Lambda$), calculated (a) from the model and (b) by RCWA. The ranges of $k_x$ and $k_y$ are in the ratio $|c|:|d|$ of the model. Colors show the polarization orientation angle $\psi$, and segments the polarization direction.}
\end{figure}

Figure~\ref{fig:vortex} shows the far-field polarization around the $\Gamma$-point BIC in the vertically symmetric lattice of Fig.~2 of the main text at $\rho=0.15$ and $0.6$, away from the charge flips. The polarization winds once around the $\Gamma$ point, in the sense of the loop at $\rho=0.15$ and against it at $\rho=0.6$, so that $q=+1$ and $-1$, respectively, in both the model and RCWA. The polarization angles of the two calculations also agree well.

\section{S3. Charge flips and Janus windows}
\subsection{Vertically symmetric lattice}
For $\epsilon_{\mathrm c}=\epsilon_{\mathrm s}$, the profile $\phi$ and the bound responses $G_m[\phi]$ and $G_m^{\TM}[\phi]$ are even in $z$, so that $\tilde S_x$ and $\tilde S_y$ are even. With $r_{\mathrm c}=r_{\mathrm s}=r$, Eq.~\eqref{eq:EF} then gives
\begin{equation}\label{eq:Fsym}
F_j^\uparrow=F_j^\downarrow=\big(1+re^{i\mu t}\big)\hat F_j,\qquad \hat F_j=\int\tilde S_j\cos\mu z\,dz,
\end{equation}
where the second term of Eq.~\eqref{eq:Fy0}, with $u^\sigma$ replaced by $\cos\mu z$, is added to $\hat F_y$, as to $F_y^\sigma$ in Eq.~\eqref{eq:EF}. To leading order in the modulation, $F_x$ and $F_y$ thus share the phase factor $1+re^{i\mu t}$ in both channels, $F_x^{\sigma*}F_y^\sigma$ is real, and Eq.~\eqref{eq:q7} reduces to Eq.~(10) of the main text,
\begin{equation}\label{eq:q8}
q^\uparrow=q^\downarrow=\sgn\big[\hat F_x\hat F_yG_{\mathrm{band}}\big].
\end{equation}
In the vertically symmetric lattice, the charge flips when one of the three real factors in Eq.~\eqref{eq:q8} changes sign, which happens in two ways. Because the two channels have the same radiation factors, they flip together in both cases.

\emph{Merging-induced charge flips.} The charge flips when $\hat F_y$ or $\hat F_x$ changes sign while $G_{\mathrm{band}}$, and hence the gap, stays finite. Apart from the common factor $1+re^{i\mu t}$, the factor $\hat F_y$ is the $k_x\to0$ limit of the radiation factor $F_y^\sigma(k_x)$ of Eq.~\eqref{eq:Fyk}, which gives the $y$-polarized far field on the $k_x$ axis per unit symmetric amplitude $A+B$. Because the far field on that axis is purely $y$ polarized, a zero of $F_y^\sigma(k_x)$ at $k_x\neq0$ is a zero of the entire far field, that is, an off-$\Gamma$ BIC. The $C_2^z$ symmetry interchanges $\tilde S_+$ and $\tilde S_-$ and the amplitudes $A$ and $B$ under $k_x\to-k_x$, so that $F_y^\sigma(k_x)$ is even, and such BICs come in pairs at $\pm k_x$. Writing $A$ and $B$ in terms of $A\pm B$ gives
\begin{equation}\label{eq:Fydec}
F_y^\sigma(k_x)=\frac12\int u^\sigma\big(\tilde S_++\tilde S_-\big)dz+\frac{A-B}{A+B}\,\frac12\int u^\sigma\big(\tilde S_+-\tilde S_-\big)dz,
\end{equation}
in which $(A-B)/(A+B)=2(h_{\mathrm g}+ih_{\mathrm r})/k_x$ diverges as $k_x\to0$, while $\tilde S_+-\tilde S_-$, which arises from the different propagation constants of the two guided waves, vanishes linearly. The limit is therefore the finite value of Eq.~\eqref{eq:Fy0}, which adds the contribution of $\tilde S_+-\tilde S_-$ to the $\Gamma$-point overlap $\int u^\sigma\tilde S_y\,dz$. Because this limit is the factor in Eq.~\eqref{eq:q8}, the sign change of $\hat F_y$ is the point at which a pair of off-$\Gamma$ BICs on the $k_x$ axis reaches the $\Gamma$ point and merges with the symmetry-protected BIC. The sign change of $\hat F_x$, the $k_y\to0$ limit of $F_x^\sigma(k_y)$ of Eq.~\eqref{eq:Exky}, corresponds in the same way to the merging of a pair of off-$\Gamma$ BICs on the $k_y$ axis.

\emph{Band-flip-induced charge flips.} The charge also flips when $G_{\mathrm{band}}$ changes sign at a band flip, where the gap closes while $\hat F_x$ and $\hat F_y$ stay finite. Near the $\Gamma$-point BIC, the model gives the $x$-polarized far field $c^\sigma k_y$ from the antisymmetric amplitude $A-B=2a$ and the $y$-polarized far field $d^\sigma k_x$ from the symmetric amplitude $A+B=ak_x/(h_{\mathrm g}+ih_{\mathrm r})$ of Eq.~\eqref{eq:ABk}. The symmetric amplitude arises because the detuning $k_x$ makes the amplitudes of the two guided waves unequal, and it is inversely proportional to $h_{\mathrm g}+ih_{\mathrm r}=Kn_{\mathrm g}(\Omega_{\mathrm a}-\Omega_{\mathrm s})/2$, that is, to the complex frequency separation of the two band-edge states. Using $F_j^\sigma=(1+re^{i\mu t})\hat F_j$ of Eq.~\eqref{eq:Fsym}, Eq.~\eqref{eq:cd} gives, for both radiation channels,
\begin{equation}\label{eq:dc}
\frac{d^\sigma}{c^\sigma}=-\frac{K\hat F_y}{\hat F_x}\,\frac{1}{G_{\mathrm{band}}+i\Rea h_{\mathrm r}}.
\end{equation}
Thus, apart from the sign associated with the direction of $\mathbf k$, the relative phase between $E_y^\sigma$ and $E_x^\sigma$ is determined by $G_{\mathrm{band}}$ and $\Rea h_{\mathrm r}$. Away from the band flip, the relative phase differs from $\pm\pi/2$, so the far field forms no lines of circular polarization. At the band flip, $G_{\mathrm{band}}=0$, the real parts of the two band-edge frequencies coincide, and their complex frequency separation becomes purely imaginary, with its magnitude set by the radiative loss of the leaky state. Consequently, $d^\sigma/c^\sigma$ becomes purely imaginary. The two far-field components are then in quadrature for every $\mathbf k$, and the polarization is circular where their amplitudes are equal, $|c^\sigma k_y|=|d^\sigma k_x|$, that is, along the two lines
\begin{equation}\label{eq:lines}
k_y=\pm\frac{K|\hat F_y|}{k_{\mathrm{EP}}|\hat F_x|}\,k_x,
\end{equation}
which cross at the $\Gamma$ point. The sign change of $G_{\mathrm{band}}$ in Eq.~\eqref{eq:q8} is therefore the point at which the two lines of circular polarization form.

\subsection{Vertically asymmetric lattice}
For $\epsilon_{\mathrm c}\neq\epsilon_{\mathrm s}$, $\phi$ and the bound responses are no longer even in $z$, and $u^\uparrow\neq u^\downarrow$ in Eq.~\eqref{eq:uK} because $r_{\mathrm c}\neq r_{\mathrm s}$, so that $F_x^\sigma$ and $F_y^\sigma$ acquire a relative phase $\theta^\sigma=\arg\big(F_x^{\sigma*}F_y^\sigma\big)$, which is in general nonzero and differs between the two channels. Equation~\eqref{eq:q7} then flips the charge of channel $\sigma$ where
\begin{equation}\label{eq:flip}
G_{\mathrm{band}}\cos\theta^\sigma+\Rea h_{\mathrm r}\sin\theta^\sigma=0,
\end{equation}
which is met in two ways. Near a band flip, $\theta^\sigma$ hardly changes, and Eq.~\eqref{eq:flip} is met as $G_{\mathrm{band}}$ passes through zero, at a small value of $G_{\mathrm{band}}$ set by the term $\Rea h_{\mathrm r}\sin\theta^\sigma$. Because $\theta^\uparrow\neq\theta^\downarrow$, this value differs between the two channels. Near a merging point, the gap stays open and $G_{\mathrm{band}}$ keeps its sign, and Eq.~\eqref{eq:flip} is met as $\theta^\sigma$ changes, because the phase of $F_y^\sigma$ or $F_x^\sigma$ turns by nearly $\pi$ there. Because the phases turn differently in the two channels, $\theta^\uparrow$ and $\theta^\downarrow$ satisfy Eq.~\eqref{eq:flip} at different values of the lattice parameter. In both cases, the BIC is therefore a Janus BIC between the flips of the two channels.

\subsection{Width of the band-flip-induced Janus window}
Near a band flip at $\rho_{\mathrm{BF}}$, dividing Eq.~\eqref{eq:flip} by $\cos\theta^\sigma$ gives $G_{\mathrm{band}}=-\tan\theta^\sigma\,\Rea h_{\mathrm r}$. We expand $G_{\mathrm{band}}\simeq G_{\mathrm{band}}'(\rho-\rho_{\mathrm{BF}})$ with $G_{\mathrm{band}}'=dG_{\mathrm{band}}/d\rho$ at $\rho_{\mathrm{BF}}$, and take $\theta^\sigma$ and $\Rea h_{\mathrm r}$, which vary slowly with $\rho$, at $\rho_{\mathrm{BF}}$. The flip of channel $\sigma$ then lies at
\begin{equation}
\rho^\sigma=\rho_{\mathrm{BF}}-\frac{\tan\theta^\sigma\,\Rea h_{\mathrm r}}{G_{\mathrm{band}}'},
\end{equation}
to first order in $\Rea h_{\mathrm r}$. The two channels carry opposite charges between $\rho^\uparrow$ and $\rho^\downarrow$, and the width of the Janus window is
\begin{equation}\label{eq:width}
\Delta\rho_{\mathrm J}=|\rho^\uparrow-\rho^\downarrow|\simeq\frac{|\tan\theta^\uparrow-\tan\theta^\downarrow|\,k_{\mathrm{EP}}}{|dG_{\mathrm{band}}/d\rho|},
\end{equation}
with $k_{\mathrm{EP}}=|\Rea h_{\mathrm r}|$ of Eq.~\eqref{eq:dispBF}, which is Eq.~(11) of the main text. The window is wide when the radiative coupling is strong, when $\tan\theta^\uparrow$ and $\tan\theta^\downarrow$ differ strongly, and when the gap closes slowly with $\rho$.

The factor $|\tan\theta^\uparrow-\tan\theta^\downarrow|$ of Eq.~\eqref{eq:width} is set by the vertical profiles of the two sources. Because, to leading order in the modulation, $F_j^\sigma=\int u^\sigma\tilde S_j\,dz$, the ratio $F_y^\sigma/F_x^\sigma$ compares the overlaps of the same radiation profile $u^\sigma$ with $\tilde S_y$ and $\tilde S_x$. If the two sources had the same vertical profile, $\tilde S_y\propto\tilde S_x$ with a real factor, this ratio would be real in both channels, and $\tan\theta^\uparrow=\tan\theta^\downarrow=0$ whatever the claddings. Within a channel, the parts of the sources that are even in $z$ enter both overlaps with the same factor, so that a nonzero $\tan\theta^\sigma$ requires the $x$- and $y$-polarized sources to differ in their vertical asymmetry. Because $u^\uparrow$ and $u^\downarrow$ weight this difference with different complex factors, $\theta^\uparrow$ and $\theta^\downarrow$ also differ, and the window opens. For the lattices of the main text, whose upper and lower interfaces are flat, both sources take their vertical asymmetry from the claddings through the guided-mode profile $\phi$ and differ little in it, so that $|\tan\theta^\uparrow-\tan\theta^\downarrow|$ remains small.

\subsection{Comparison with RCWA}
\begin{figure}[p]
\includegraphics[scale=0.87]{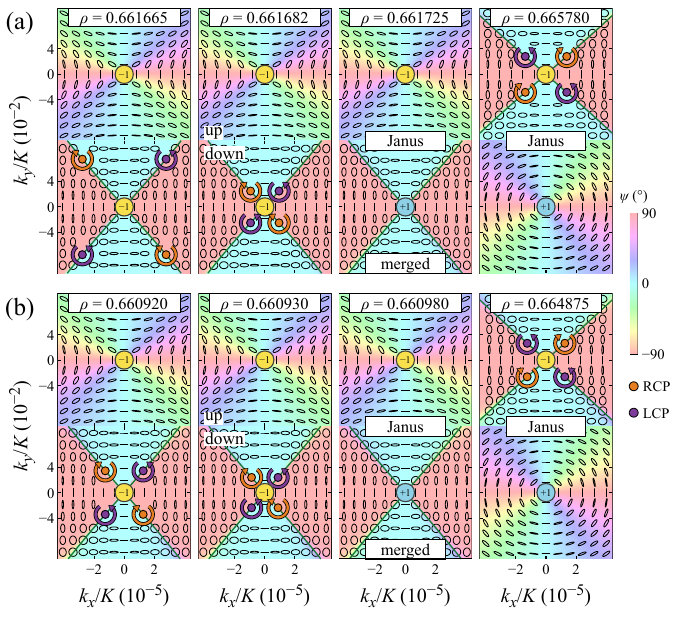}
\caption{\label{fig:cmp3}Far-field polarization maps of Fig.~3(b) of the main text ($\epsilon_{\mathrm c}=1$, $\epsilon_{\mathrm s}=2.1316$, $\epsilon_{\mathrm{avg}}=12$, $\Delta\epsilon=1$, $t=0.5\Lambda$), calculated (a) from the model and (b) by RCWA. Colors show the polarization orientation angle $\psi$, and green lines mark the directions in which $|E_x^{\sigma}|=|E_y^{\sigma}|$ near each channel flip.}
\end{figure}

\begin{figure}[p]
\includegraphics[scale=0.87]{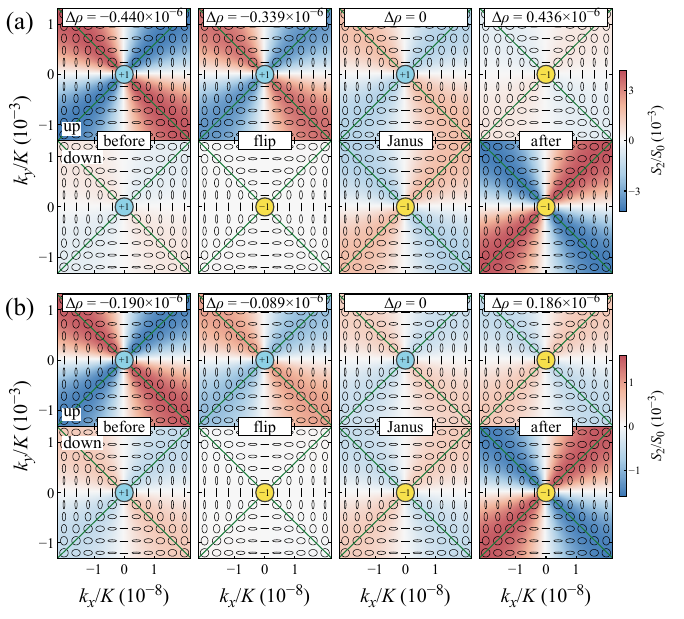}
\caption{\label{fig:cmp4}Far-field polarization maps of Fig.~4(c) of the main text ($\epsilon_{\mathrm c}=1$, $\epsilon_{\mathrm s}=2.1316$, $\epsilon_{\mathrm{avg}}=12$, $\Delta\epsilon=1$, $t=0.5\Lambda$), calculated (a) from the model and (b) by RCWA, overlaid on $S_2/S_0$. Green lines mark the directions in which $|E_x^{\sigma}|=|E_y^{\sigma}|$. $\Delta\rho$ is measured from the center of each window.}
\end{figure}

Figures~\ref{fig:cmp3} and \ref{fig:cmp4} compare the far-field maps of Figs.~3(b) and 4(c) of the main text, calculated from the model and by RCWA. In Fig.~\ref{fig:cmp3}, the RCWA maps of the first and third columns are calculated at the same distances from the downward-channel flip as the model maps, $1.92\times10^{-5}$ below and $4.08\times10^{-5}$ above it. At equal distances from the flips, RCWA places the circularly polarized points (C points) 1.8--2.2 times closer to $\Gamma$. The RCWA maps of the second and fourth columns are therefore calculated $9.2\times10^{-6}$ below the downward-channel flip and $1.94\times10^{-5}$ below the upward-channel flip, where the C points lie about as far from $\Gamma$ as in the model maps. The charges of both channels agree between the two calculations in all four columns, and the C points lie along the same directions in the two calculations.

In Fig.~\ref{fig:cmp4}, the Janus window is $1.77\times10^{-7}$ wide in RCWA and $6.8\times10^{-7}$ wide in the model. The RCWA maps are therefore calculated at the positions of the model maps relative to the channel flips, $1.0\times10^{-7}$ below the downward-channel flip, at the downward-channel flip, midway between the two flips, and $1.0\times10^{-7}$ above the upward-channel flip. The charges of both channels agree between the two calculations in all four columns. At the downward-channel flip, $|S_2/S_0|$ of the downward channel stays close to zero away from $\Gamma$ in both calculations, and the downward far field forms lines of circular polarization along the directions in which $|E_x^{\downarrow}|=|E_y^{\downarrow}|$.

\section{S4. Second stop band}
The second stop band is formed by the counterpropagating TE$_0$ guided waves at $\pm K$, and the derivation of Secs.~S1--S3 applies with the changes given below. In this section, $\phi$, $\Omega_0$, $n_{\mathrm g}$, $h_{\mathrm g}$, $h_{\mathrm r}$, $\tilde S_y$, and $\tilde S_x$ denote the corresponding quantities of the second stop band.

\subsection{Field approximation and coupling paths}
Retaining Fourier components through $\epsilon_2$, we approximate the electric field as
\begin{equation}\label{eq:field2}
\mathbf E=\Big[\big(A\hat{\mathbf e}_+e^{iKx}+B\hat{\mathbf e}_-e^{-iKx}\big)\phi+\mathbf E_{\mathrm{rad}}+\mathbf E_{+2}e^{i2Kx}+\mathbf E_{-2}e^{-i2Kx}\Big]e^{i(k_xx+k_yy)},
\end{equation}
where $\ex_\pm=\ex_{\pm1}\simeq(\mp k_y/K,1)$ and $\phi$ is the transverse profile of the TE$_0$ mode of the reference slab, whose propagation constant equals $K$ at the Bragg frequency $\Omega_0$. Through $\epsilon_1$, each guided wave radiates into $\mathbf E_{\mathrm{rad}}$ ($+K\to0$) and excites the intermediate harmonic on its own side ($+K\to+2K$). Through $\epsilon_2$, the two guided waves couple to each other ($+K\to-K$), and the intermediate harmonics radiate into $\mathbf E_{\mathrm{rad}}$ ($+2K\to0$). The intermediate harmonics are thus driven by $\mathbf P_{\pm2}=\epsilon_1\mathbf E_{\pm1}$, and Eq.~\eqref{eq:resp} gives
\begin{equation}\label{eq:E2nd}
\mathbf E_{\pm2}=-\frac{k_0^2}{2}\epsilon_1A_\pm\Big[\ex_{\pm2}(\ex_{\pm2}\cdot\ex_\pm)G_{\pm2}+\px_{\pm2}(\px_{\pm2}\cdot\ex_\pm)G_{\pm2}^{\TM}\Big][\phi].
\end{equation}
The TE part of $\epsilon_2\mathbf E_{\mp1}$ couples the guided waves to each other, and its TM part gives the bound TM field of the $\pm K$ harmonic,
\begin{equation}\label{eq:E1p}
\mathbf E_{\pm1,p}=-\frac{k_0^2}{2}\epsilon_2A_\mp(\px_{\pm1}\cdot\ex_\mp)\,\px_{\pm1}\,G_{\pm1}^{\TM}[\phi].
\end{equation}
The radiating harmonic is driven by
\begin{equation}\label{eq:P02}
\mathbf P_0=\epsilon_1\big(\mathbf E_1+\mathbf E_{-1}\big)+\epsilon_2\big(\mathbf E_2+\mathbf E_{-2}\big),
\end{equation}
where $\mathbf E_{\pm1}=A_\pm\ex_\pm\phi+\mathbf E_{\pm1,p}$.

\subsection{Coupled-mode equations and dispersion relation}
Projecting the TE parts of the $m=\pm1$ equations on $\phi$ as in Sec.~S1, with $k_0^2-K^2\Omega_0^2\simeq2K^2\Omega_0(\Omega-\Omega_0)$ and $|\mathbf q_{\pm1}|^2-K^2\simeq\pm2Kk_x$, and dividing by $-2K\int\phi^2dz$, gives Eq.~(3) of the main text with $h_{AA}=ih_{\mathrm r}$, $h_{AB}=h_{\mathrm g}+ih_{\mathrm r}$, and
\begin{align}
n_{\mathrm g}&=\frac{\Omega_0\int\epsilon_0\phi^2dz}{\int\phi^2dz},\qquad N_\phi=2K\!\int\!\phi^2dz,\\
h_{\mathrm g}&=\frac{k_0^2}{2N_\phi}\Big(\epsilon_2\!\int\!\phi^2dz+\frac{k_0^2}{2}\epsilon_1^2\!\int\!\phi\,G_2[\phi]\,dz\Big),\label{eq:hg2}\\
h_{\mathrm r}&=\frac{ik_0^4}{4N_\phi}\Big(\int\tilde S_y\,G_0[\tilde S_y]\,dz+\epsilon_1^2\!\int\!\phi\,G_2[\phi]\,dz\Big),\qquad
\tilde S_y=\epsilon_1\phi-\frac{k_0^2}{2}\epsilon_1\epsilon_2G_2[\phi],\label{eq:hr2}
\end{align}
where $\mathbf P_0=(A+B)\tilde S_y\hat{\mathbf y}$ at the $\Gamma$ point. The $\epsilon_1^2G_2$ terms come from the bound paths $\pm K\to\pm2K\to\pm K$, which return each guided wave to itself. The dispersion relation is
\begin{equation}\label{eq:disp2}
\Omega(k_x)=\Omega_0-\frac{ih_{\mathrm r}\pm\sqrt{k_x^2+(h_{\mathrm g}+ih_{\mathrm r})^2}}{Kn_{\mathrm g}},
\end{equation}
with the BIC at $\Omega_{\mathrm a}=\Omega_0+h_{\mathrm g}/(Kn_{\mathrm g})$, the leaky state at $\Omega_{\mathrm s}=\Omega_0-(h_{\mathrm g}+2ih_{\mathrm r})/(Kn_{\mathrm g})$, and the band gap $2|G_{\mathrm{band}}|/(Kn_{\mathrm g})$ with $G_{\mathrm{band}}=h_{\mathrm g}-\Ima h_{\mathrm r}$. The BIC branch near the $\Gamma$ point again has $A+B=ak_x/(h_{\mathrm g}+ih_{\mathrm r})$ and $A-B=2a$, and $k_{\mathrm{EP}}=|\Rea h_{\mathrm r}|$.

\subsection{Radiation source and charge}
To first order in $k_y$, $\hat e_{m,x}\simeq-k_y/mK$, $\hat p_{m,x}\simeq1$, and $\px_m\cdot\ex_\pm\simeq(k_y/K)(1/m\mp1)$. Inserting these into Eqs.~\eqref{eq:E2nd}--\eqref{eq:P02} gives the two terms of $P_{0,x}$,
\begin{align}
\epsilon_1\big(E_{1,x}+E_{-1,x}\big)&=-(A-B)\frac{k_y}{K}\,\epsilon_1\big(\phi-k_0^2\epsilon_2G_1^{\TM}[\phi]\big),\nonumber\\
\epsilon_2\big(E_{2,x}+E_{-2,x}\big)&=(A-B)\frac{k_y}{K}\,\frac{k_0^2}{2}\epsilon_1\epsilon_2\,\tfrac12\big(G_2+G_2^{\TM}\big)[\phi],
\end{align}
so that the source near the $\Gamma$ point is $\mathbf S=(A+B)\tilde S_y\hat{\mathbf y}-(A-B)(k_y/K)\tilde S_x\hat{\mathbf x}$ with $\tilde S_y$ of Eq.~\eqref{eq:hr2} and
\begin{equation}\label{eq:Sx2}
\tilde S_x=\epsilon_1\big(\phi-k_0^2\epsilon_2G_1^{\TM}[\phi]\big)-\frac{k_0^2}{2}\epsilon_1\epsilon_2\,\tfrac12\big(G_2+G_2^{\TM}\big)[\phi].
\end{equation}
The term $-k_0^2\epsilon_1\epsilon_2G_1^{\TM}[\phi]$ of $\tilde S_x$ comes from the TM part of the $\pm K$ harmonics, Eq.~\eqref{eq:E1p}. With the radiation profiles and coupling constants of Eq.~\eqref{eq:uK}, the far field is $\mathbf E_{\mathrm{rad}}^\sigma=(c^\sigma k_y,\,d^\sigma k_x)$ with
\begin{equation}
c^\sigma=\frac{k_0^2}{K}\,\mathcal K^\sigma a\,F_x^\sigma,\qquad d^\sigma=-\frac{k_0^2}{2}\,\frac{\mathcal K^\sigma a}{h_{\mathrm g}+ih_{\mathrm r}}\,F_y^\sigma,\qquad F_j^\sigma=\int u^\sigma\tilde S_j\,dz,
\end{equation}
where, as in Eq.~\eqref{eq:EF}, the second term of Eq.~\eqref{eq:Fy0}, formed with the sources of the two guided waves at $\pm K$, is added to $F_y^\sigma$. The coefficient $c^\sigma$ differs from that of Eq.~\eqref{eq:cd} only by the positive factor $2$, which comes from the tilt $\mp k_y/K$ of $\ex_\pm$. Hence $c^{\sigma*}d^\sigma=-(k_0^4/2K)|\mathcal K^\sigma a|^2F_x^{\sigma*}F_y^\sigma/(h_{\mathrm g}+ih_{\mathrm r})$ has the same sign structure as Eq.~\eqref{eq:qF}, so that Eqs.~(9) and (10) of the main text and the analysis of Sec.~S3 hold unchanged.

\subsection{Charge flips}
In the symmetric lattice, the charge is set by the signs of $G_{\mathrm{band}}$, $\hat F_y$, and $\hat F_x$ in Eq.~\eqref{eq:q8}. At the second stop band, their leading terms are set by different Fourier components. $G_{\mathrm{band}}$ follows the first-order term $(k_0^2/2N_\phi)\epsilon_2\int\phi^2dz$ of $h_{\mathrm g}$ in Eq.~\eqref{eq:hg2}, and $\hat F_y$ and $\hat F_x$ follow $\epsilon_1\int\phi\cos\mu z\,dz$, because the bound responses in $\tilde S_y$ and $\tilde S_x$ are of higher order in the modulation. Varying $\rho$ changes the Fourier components. Because $\epsilon_2=(\Delta\epsilon/\pi)\sin2\pi\rho$ changes sign at $\rho=1/2$, $G_{\mathrm{band}}$ changes sign near $\rho=1/2$, and the charge flips at a band flip. Because $\epsilon_1=(2\Delta\epsilon/\pi)\sin\pi\rho$ has no zero for $0<\rho<1$ and the overlap, set by the reference slab, does not depend on $\rho$, $\hat F_y$ and $\hat F_x$ keep their signs. Varying $t$ changes the overlap instead. At fixed $\rho$ away from $1/2$, $G_{\mathrm{band}}$ keeps its sign, because $\epsilon_2$ is fixed and $\int\phi^2dz$ is positive, whereas $\int\phi\cos\mu z\,dz$ changes sign as $t$ varies, because the guided mode $\phi$ and the radiation profile $\cos\mu z$ have different transverse wave numbers in the layer. $\hat F_y$ and $\hat F_x$ then pass through zero at nearby thicknesses. At each zero, a pair of off-$\Gamma$ BICs, on the $k_x$ axis for $\hat F_y=0$ and on the $k_y$ axis for $\hat F_x=0$, merges with the $\Gamma$-point BIC and reverses its charge. In the vertically asymmetric lattice, both types of flips split between the two radiation channels as in Sec.~S3, so that varying $\rho$ opens a band-flip-induced Janus window and varying $t$ opens merging-induced Janus windows. Figure~5 of the main text shows these flips for $\rho$ varied at $t=0.5\Lambda$ and for $t$ varied at $\rho=0.30$.

\begin{figure}[t]
\includegraphics{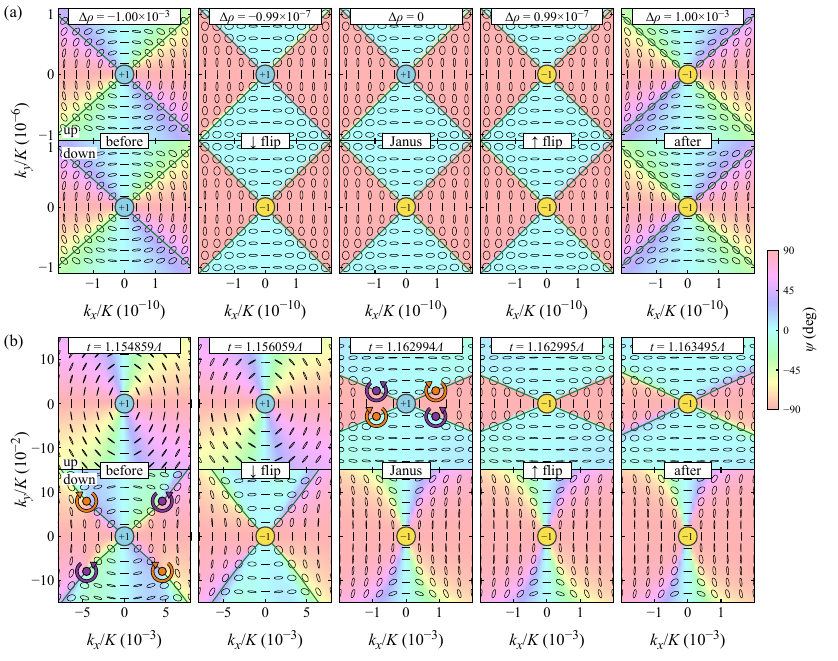}
\caption{\label{fig:2nd}Far-field polarization maps at the second stop band of the vertically asymmetric lattice ($\epsilon_{\mathrm c}=1$, $\epsilon_{\mathrm s}=2.1316$, $\epsilon_{\mathrm{avg}}=12$, $\Delta\epsilon=1$), calculated from the model, (a) across the band-flip-induced Janus window of Fig.~5(e) of the main text ($t=0.5\Lambda$) and (b) across the first merging-induced Janus window of Fig.~5(f) ($\rho=0.30$). In (a), $\Delta\rho$ is measured from the center of the window. Colors show the polarization orientation angle $\psi$, and green lines mark the directions in which $|E_x^{\sigma}|=|E_y^{\sigma}|$.}
\end{figure}

Figure~\ref{fig:2nd} shows the far-field polarization maps across the band-flip-induced Janus window and the first merging-induced Janus window. In Fig.~\ref{fig:2nd}(a), the two channels flip at the two edges of the band-flip-induced window, and at its flip the far field of each channel is circularly polarized along the lines $|E_x^{\sigma}|=|E_y^{\sigma}|$, with $|S_2/S_0|<10^{-7}$. The window is so narrow that the maps hardly change across it. Outside the window, the far fields along these lines are elliptically polarized, and the two channels carry the same charge. In Fig.~\ref{fig:2nd}(b), four C points of the downward channel lie on the lines $|E_x^{\downarrow}|=|E_y^{\downarrow}|$ before the downward-channel flip and merge with the $\Gamma$-point BIC at that flip, reversing the downward charge to $-1$. Just before the upward-channel flip, four C points of the upward channel lie close to $\Gamma$ on the lines $|E_x^{\uparrow}|=|E_y^{\uparrow}|$ and merge with the BIC at that flip. After the merging-induced Janus window, both channels carry $-1$.

\section{S5. Two-dimensional lattices}
In a square lattice of rectangular pillars, $\rho_x\Lambda$ wide and $\rho_y\Lambda$ long, the fourth stop band at the $\Gamma$ point is formed by four TE$_0$ guided waves, the $x$ pair at $(\pm2K,0)$ with amplitudes $(A,B)$ and the $y$ pair at $(0,\pm2K)$ with amplitudes $(C,D)$. We first classify the bands along the two axes and the four band-edge modes at the $\Gamma$ point, and then show how the charge flips of Secs.~S2 and S3 extend to two dimensions and why rectangular rather than square pillars are needed.

\subsection{Four-wave coupled-mode model}
Each pair alone obeys the coupled-mode equations of Sec.~S1, with its own couplings $h_{\mathrm g}^{x,y}$ and $h_{\mathrm r}^{x,y}$ and the detuning $k_x$ or $k_y$. The two pairs are coupled through the intermediate harmonics, and the four-wave system reads
\begin{equation}\label{eq:M4}
Kn_{\mathrm g}(\Omega_0-\Omega)\begin{pmatrix}A\\B\\C\\D\end{pmatrix}=\begin{pmatrix}H_x(k_x)&V\\V^{\mathrm T}&H_y(k_y)\end{pmatrix}\begin{pmatrix}A\\B\\C\\D\end{pmatrix},
\end{equation}
where $H_x(k_x)$ is the matrix of Eq.~(3) of the main text for the $x$ pair, with $h_{AA}=ih_{\mathrm r}^x$ and $h_{AB}=h_{\mathrm g}^x+ih_{\mathrm r}^x$, $H_y(k_y)$ the same matrix for the $y$ pair with $h_{\mathrm g}^y$, $h_{\mathrm r}^y$, and $k_y$, and $V$ the coupling between the pairs through the intermediate harmonics, evaluated at the $\Gamma$ point, which connects each wave of one pair with both waves of the other. As in Sec.~S1, $h_{\mathrm g}^{x,y}$ and $h_{\mathrm r}^{x,y}$ include the bound paths back to the same guided wave, which differ between the two pairs for $\rho_x\neq\rho_y$.

Inside the grating layer, the rectangular pillars give
\begin{equation}\label{eq:eps2d}
\epsilon(x,y,z)=\epsilon_0+\frac12\sum_{\mathbf m\neq\mathbf 0}\epsilon_{\mathbf m}e^{i(mx+ny)K},\qquad \epsilon_{\mathbf m}=2\Delta\epsilon\,s_m(\rho_x)\,s_n(\rho_y),
\end{equation}
for $\mathbf m=(m,n)$, with $s_0(\rho)=\rho$ and $s_n(\rho)=\sin(n\pi\rho)/(n\pi)$, so that $\epsilon_{(m,0)}$ equals $\epsilon_m$ of Sec.~S1 for $\rho_y=1$. Besides the guided waves at $\mathbf m_I=(\pm2,0)$ and $(0,\pm2)$ for $I=A,B,C,D$ and the radiating harmonic $\mathbf E_{\mathrm{rad}}$, the field contains all harmonics $\mathbf m$ with $|m|,|n|\le7$ at the in-plane wave vectors $\mathbf k+\mathbf mK$, each driven to first order by the guided waves as in Eq.~\eqref{eq:resp}. At $\mathbf m=\mathbf m_I$, only the TM part is kept, because the TE part is the guided wave itself, so that $G_{\mathbf m_I}=0$ below. Projecting on $\phi$ as in Sec.~S1 gives the elements of Eq.~\eqref{eq:M4},
\begin{multline}\label{eq:hIJ}
h_{IJ}=\frac{k_0^2}{2N_\phi}\bigg[\epsilon_{\mathbf m_I-\mathbf m_J}(\ex_I\cdot\ex_J)\!\int\!\phi^2dz-\frac{k_0^2}{2}\!\int\!\tilde{\mathbf S}_I\cdot G_0[\tilde{\mathbf S}_J]\,dz\\
-\frac{k_0^2}{2}\sum_{\mathbf m}\epsilon_{\mathbf m-\mathbf m_I}\epsilon_{\mathbf m-\mathbf m_J}\!\int\!\phi\,\Big[(\ex_I\cdot\ex_J-\nu_I\nu_J)G_{\mathbf m}+\nu_I\nu_JG_{\mathbf m}^{\TM}\Big][\phi]\,dz\bigg],
\end{multline}
where $\nu_I=\hat{\mathbf q}_{\mathbf m}\cdot\ex_I$ with the unit vector $\hat{\mathbf q}_{\mathbf m}$ along $\mathbf k+\mathbf mK$, the coefficient $\epsilon_{\mathbf m}$ is taken as zero for $\mathbf m=\mathbf 0$, $N_\phi$ is that of Eq.~\eqref{eq:h0}, and
\begin{equation}\label{eq:S2d}
\tilde{\mathbf S}_J=\epsilon_{\mathbf m_J}\ex_J\phi-\frac{k_0^2}{2}\sum_{\mathbf m}\epsilon_{\mathbf m}\epsilon_{\mathbf m-\mathbf m_J}\Big[(\ex_J-\nu_J\hat{\mathbf q}_{\mathbf m})G_{\mathbf m}+\nu_J\hat{\mathbf q}_{\mathbf m}G_{\mathbf m}^{\TM}\Big][\phi]
\end{equation}
is the source of $\mathbf E_{\mathrm{rad}}$ produced by wave $J$ of unit amplitude. The elements within the $x$ and $y$ pairs form $H_x$ and $H_y$, and those between the pairs form $V$. Between the pairs, $\ex_I\cdot\ex_J=0$, because the $x$ pair carries $E_y$ and the $y$ pair carries $E_x$, and at the $\Gamma$ point $\tilde{\mathbf S}_I\cdot G_0[\tilde{\mathbf S}_J]=0$ as well. The coupling $V$ therefore comes only from the harmonics whose in-plane wave vectors lie along neither axis, through the difference between their TM and TE responses. The far field of channel $\sigma$ is Eq.~\eqref{eq:Esig} with the source $A\tilde{\mathbf S}_A+B\tilde{\mathbf S}_B+C\tilde{\mathbf S}_C+D\tilde{\mathbf S}_D$ of the eigenvector of Eq.~\eqref{eq:M4}.

\subsection{Bands along the $k_x$ and $k_y$ axes}
The mirror symmetries $x\to-x$ and $y\to-y$ of the lattice separate the symmetric and antisymmetric combinations of each pair, $X_{\mathrm s}$ and $X_{\mathrm a}$ with $A=\pm B$ and $Y_{\mathrm s}$ and $Y_{\mathrm a}$ with $C=\pm D$. Because the $x$ pair carries $E_y$ and the $y$ pair carries $E_x$, $X_{\mathrm s}$ is even under $x\to-x$ and odd under $y\to-y$, $Y_{\mathrm s}$ is odd under $x\to-x$ and even under $y\to-y$, and $X_{\mathrm a}$ and $Y_{\mathrm a}$ are odd under both. Along the $k_x$ axis, the mirror $y\to-y$ is kept, under which $X_{\mathrm s}$, $X_{\mathrm a}$, and $Y_{\mathrm a}$ are odd and $Y_{\mathrm s}$ is even. The detuning $k_x$ couples $X_{\mathrm a}$ to $X_{\mathrm s}$, as in Eq.~\eqref{eq:ABk}, and $V$ couples $X_{\mathrm a}$ to $Y_{\mathrm a}$, with $\langle Y_{\mathrm a}|V|X_{\mathrm a}\rangle=v$. The three odd states thus form one block that radiates $y$-polarized light, whereas $Y_{\mathrm s}$ is decoupled from them, radiates $x$-polarized light, and is not detuned to first order in $k_x$. With the coefficients evaluated at the $\Gamma$ point, the determinant of the block gives the dispersion relation
\begin{equation}\label{eq:disp2d}
\big[(\lambda-\lambda_{X_{\mathrm s}})(\lambda+h_{\mathrm g}^x)-k_x^2\big](\lambda+h_{\mathrm g}^y)=v^2(\lambda-\lambda_{X_{\mathrm s}}),
\end{equation}
with $\Omega=\Omega_0-\lambda/(Kn_{\mathrm g})$, where $\lambda_{X_{\mathrm s}}=h_{\mathrm g}^x+2ih_{\mathrm r}^x$ is the eigenvalue of $X_{\mathrm s}$, as in Sec.~S1. For $v=0$, the first factor gives Eq.~\eqref{eq:disp} for the $x$ pair and the second the flat level $\lambda=-h_{\mathrm g}^y$ of $Y_{\mathrm a}$. The coupling $v$ makes an $x$-pair band and the $Y_{\mathrm a}$ level repel where they meet and exchange their $x$- and $y$-pair weights, so that a band that starts from a BIC at the $\Gamma$ point can continue as a leaky band of the $x$ pair, and a band that starts from a leaky state as a band of high $Q$, as in Fig.~\ref{fig:2d}(b). Along the $k_y$ axis, the mirror $x\to-x$ is kept, $X_{\mathrm a}$, $Y_{\mathrm a}$, and $Y_{\mathrm s}$ form one block that radiates $x$-polarized light, $X_{\mathrm s}$ is decoupled from them, and the dispersion relation holds with $x$ and $y$ interchanged.

\subsection{Band-edge modes at the $\Gamma$ point}
At the $\Gamma$ point, the four states fall into three irreducible representations of $C_{2v}$. $X_{\mathrm s}$ and $Y_{\mathrm s}$ have the parities of $y$- and $x$-polarized normal emission, respectively, whereas $X_{\mathrm a}$ and $Y_{\mathrm a}$ share the third representation, which no normally emitted wave has. The leaky states therefore couple to no other state, and each has the eigenvalue of the symmetric combination of its own pair,
\begin{equation}\label{eq:lamS}
\lambda_{X_{\mathrm s}}=h_{\mathrm g}^x+2ih_{\mathrm r}^x,\qquad \lambda_{Y_{\mathrm s}}=h_{\mathrm g}^y+2ih_{\mathrm r}^y,
\end{equation}
whereas $X_{\mathrm a}$ and $Y_{\mathrm a}$, with the real eigenvalues $-h_{\mathrm g}^x$ and $-h_{\mathrm g}^y$, are coupled by $v$. Equation~\eqref{eq:disp2d} at $k_x=0$ accordingly factorizes into $\lambda=\lambda_{X_{\mathrm s}}$ and $(\lambda+h_{\mathrm g}^x)(\lambda+h_{\mathrm g}^y)=v^2$. The upper and lower $\Gamma$-point BICs are therefore
\begin{equation}\label{eq:XY}
\begin{gathered}
|U\rangle=\cos\xi\,|X_{\mathrm a}\rangle+\sin\xi\,|Y_{\mathrm a}\rangle,\qquad
|L\rangle=-\sin\xi\,|X_{\mathrm a}\rangle+\cos\xi\,|Y_{\mathrm a}\rangle,\\
\tan\xi=-\frac{v}{(h_{\mathrm g}^x-h_{\mathrm g}^y)/2+R},\qquad R=\big[(h_{\mathrm g}^x-h_{\mathrm g}^y)^2/4+v^2\big]^{1/2},
\end{gathered}
\end{equation}
with $|\xi|<\pi/2$ and the real eigenvalues $\lambda_{U,L}=-(h_{\mathrm g}^x+h_{\mathrm g}^y)/2\mp R$. For $\rho_y\to1$, the modulation along $y$ vanishes, $v\to0$, and the $x$ pair obeys the coupled-mode equations of the 1D lattice of Secs.~S1--S3. When $h_{\mathrm g}^x>h_{\mathrm g}^y$ in this limit, as for $\rho_x=0.55$, $\xi\to0$ and $|U\rangle$ becomes the BIC of the 1D lattice.

The four levels move with $\rho_y$ through the Fourier components that set the couplings $h_{\mathrm g}^x$ and $h_{\mathrm g}^y$ within the two pairs. Within the $x$ pair, the first-order coupling is $\epsilon_{(4,0)}=2\Delta\epsilon\,\rho_y\,s_4(\rho_x)$, which grows linearly with $\rho_y$, so that the $x$-pair levels change slowly. Within the $y$ pair, it is $\epsilon_{(0,4)}=2\Delta\epsilon\,\rho_x\,s_4(\rho_y)$, which changes sign at $\rho_y=1/4$, $1/2$, and $3/4$, so that the $y$-pair levels oscillate, and $Y_{\mathrm s}$ and $Y_{\mathrm a}$ exchange their order near these points, as at the band flips of Sec.~S3. Varying $\rho_x$ at fixed $\rho_y$ gives the same behavior with the $x$ and $y$ pairs interchanged, because a rotation by $90^\circ$ maps the lattice with $(\rho_x,\rho_y)$ onto that with $(\rho_y,\rho_x)$. The two BICs share one representation and therefore do not cross as $\rho_y$ varies, unless $v$ vanishes at $\rho_y=\rho_x$. $|U\rangle$ is $x$-like ($|\xi|<\pi/4$) where $h_{\mathrm g}^x>h_{\mathrm g}^y$ and $y$-like ($|\xi|>\pi/4$) where $h_{\mathrm g}^x<h_{\mathrm g}^y$, and $|L\rangle$ is the reverse. Where $h_{\mathrm g}^x=h_{\mathrm g}^y$, they are separated by $2|v|$ and $|\xi|$ passes through $\pi/4$, so that $|U\rangle$ and $|L\rangle$ exchange their $x$ and $y$ character. The exchange occurs at $\rho_y=\rho_x$, where the $C_{4v}$ symmetry makes $h_{\mathrm g}^x=h_{\mathrm g}^y$ and the leaky states $X_{\mathrm s}$ and $Y_{\mathrm s}$ degenerate, and wherever the two couplings coincide accidentally. The leaky states can cross each other and either BIC without level repulsion, and a crossing of either BIC with $X_{\mathrm s}$ or $Y_{\mathrm s}$ produces the first two kinds of charge flips below.

\begin{figure}[t]
\includegraphics{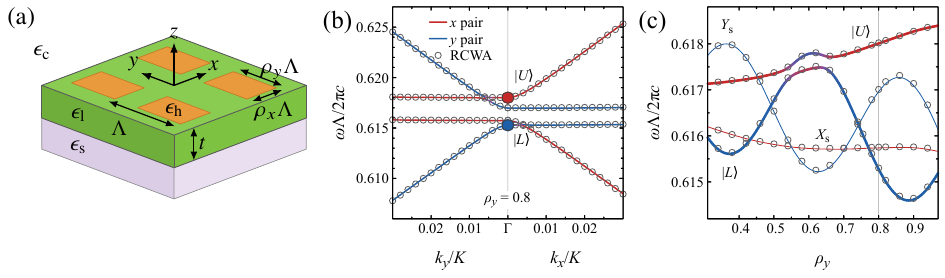}
\caption{\label{fig:2d}(a) Schematic of the 2D lattice of rectangular pillars. (b), (c) Results for $\epsilon_{\mathrm c}=1$, $\epsilon_{\mathrm s}=2.1316$, $\epsilon_{\mathrm{avg}}=12$, $\Delta\epsilon=1$, $t=0.5\Lambda$, and $\rho_x=0.55$, calculated from the model and by RCWA. (b) Bands near the $\Gamma$ point at $\rho_y=0.8$, marked by the gray line in (c). (c) Frequencies of the four $\Gamma$-point states as functions of $\rho_y$, with thick lines for the BICs $|U\rangle$ and $|L\rangle$ and thin lines for the leaky states $X_{\mathrm s}$ and $Y_{\mathrm s}$. The model lines are colored by the weight of the $x$ pair, from blue for the $y$ pair to red for the $x$ pair.}
\end{figure}

Figure~\ref{fig:2d}(c) shows the four $\Gamma$-point states for $\rho_x=0.55$. The $x$-pair levels change slowly with $\rho_y$, whereas the $y$-pair levels oscillate, and $|L\rangle$, which is $y$-like there, crosses $X_{\mathrm s}$ at $\rho_y=0.43$ and $0.77$ and $Y_{\mathrm s}$ at $0.49$ and $0.74$. The colors of $|U\rangle$ and $|L\rangle$ change at $\rho_y=0.55$ and $0.645$, where the two BICs repel each other and exchange their $x$ and $y$ character, and the leaky states cross at $\rho_y=0.55$ and $0.70$. $Y_{\mathrm s}$ crosses $|U\rangle$ at $\rho_y=0.45$, which gives the second kind of charge flip below, whereas $|U\rangle$ does not cross $X_{\mathrm s}$ in this range, because the $x$-pair gap keeps its sign. At $\rho_y=0.8$, the states lie in the order $|U\rangle$, $Y_{\mathrm s}$, $X_{\mathrm s}$, and $|L\rangle$ from top to bottom, as in Fig.~\ref{fig:2d}(b). The model results are calculated as described in Sec.~S6, with the frequency iterated to the eigenfrequency of each band. The two calculations agree well in frequency along both axes at $\rho_y=0.8$ and at the $\Gamma$ point for $0.30\le\rho_y\le0.98$.

\subsection{Far field and charge flips}
Away from the $\Gamma$ point, the detunings make the amplitudes of each pair unequal, as in Eq.~\eqref{eq:ABk}. Along $k_x$, the $X_{\mathrm a}$ component $\cos\xi$ of $|U\rangle$ acquires $A+B\propto k_x\cos\xi/(\lambda_U-\lambda_{X_{\mathrm s}})$, which radiates $E_y$, and along $k_y$, its $Y_{\mathrm a}$ component $\sin\xi$ acquires $C+D\propto k_y\sin\xi/(\lambda_U-\lambda_{Y_{\mathrm s}})$, which radiates $E_x$. To first order in $\mathbf k$, the far field keeps the form $\mathbf E_{\mathrm{rad}}^\sigma=(c^\sigma k_y,\,d^\sigma k_x)$ with
\begin{equation}\label{eq:cd2d}
c^\sigma=c_0^\sigma-\frac{\sin\xi\,E_{Y_{\mathrm s}}^\sigma}{\lambda_U-\lambda_{Y_{\mathrm s}}},\qquad
d^\sigma=d_0^\sigma-\frac{\cos\xi\,E_{X_{\mathrm s}}^\sigma}{\lambda_U-\lambda_{X_{\mathrm s}}},
\end{equation}
where $E_{X_{\mathrm s}}^\sigma$ and $E_{Y_{\mathrm s}}^\sigma$ are the amplitudes radiated by $X_{\mathrm s}$ and $Y_{\mathrm s}$ into channel $\sigma$, and $c_0^\sigma$ and $d_0^\sigma$ collect the remaining first-order terms without a pole, such as the radiation from the tilt of the polarization vectors in Eq.~\eqref{eq:cd}. For $\xi\to0$, $\lambda_U-\lambda_{X_{\mathrm s}}\to-2(h_{\mathrm g}^x+ih_{\mathrm r}^x)$, and $c^\sigma$ and $d^\sigma$ take the form of Eq.~\eqref{eq:cd}. The charge is again $q^\sigma=-\sgn\Rea(c^{\sigma*}d^\sigma)$, and, as $\rho_y$ is varied, it flips in three ways.
\begin{enumerate}
\item When $|U\rangle$ crosses $X_{\mathrm s}$, $\Rea(\lambda_U-\lambda_{X_{\mathrm s}})$ passes through zero, and the real part of the second term of $d^\sigma$ in Eq.~\eqref{eq:cd2d} changes sign. For $\xi\to0$, this is the band flip of Sec.~S3.
\item At the crossing of $|U\rangle$ with $Y_{\mathrm s}$, $\Rea(\lambda_U-\lambda_{Y_{\mathrm s}})$ passes through zero, and the real part of the second term of $c^\sigma$ in Eq.~\eqref{eq:cd2d} changes sign. Where $|U\rangle$ is $x$-like, this flip has no 1D counterpart. As at the band flip, the pole, set by $\lambda_U-\lambda_{Y_{\mathrm s}}$, is the same in both channels, whereas the zero of $\Rea(c^{\sigma*}d^\sigma)$ also depends on the channel-dependent amplitudes. The flips of the two channels therefore lie close to the crossing without coinciding exactly with it or with each other, and their separation is set by $\Ima\lambda_{Y_{\mathrm s}}$ and the channel phases.
\item Away from both poles, $\Rea(c^{\sigma*}d^\sigma)$ changes sign through the interference of the terms in Eq.~\eqref{eq:cd2d}, and off-$\Gamma$ BICs or C points merge with the BIC while the gaps remain open. In a vertically asymmetric lattice, $c^\sigma$ and $d^\sigma$ have channel-dependent phases, so that this flip occurs at different $\rho_y$ in the two channels and opens Janus windows, as near $\rho_{\mathrm{F3}}$ and $\rho_{\mathrm{F4}}$ in 1D.
\end{enumerate}

\begin{figure}[t]
\includegraphics{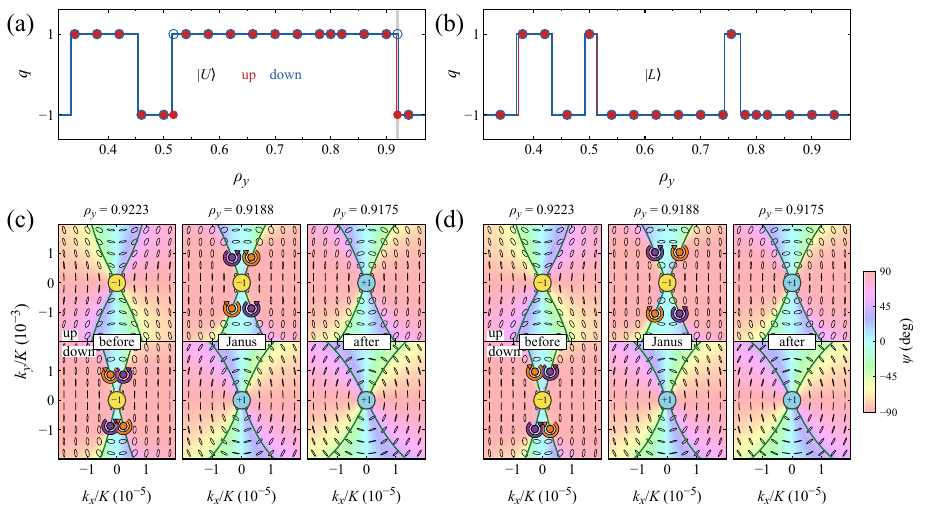}
\caption{\label{fig:2dq}(a), (b) Channel charges of (a) $|U\rangle$ and (b) $|L\rangle$ of the lattice of Fig.~\ref{fig:2d} as functions of $\rho_y$, calculated from the model (lines) and by RCWA (circles). The gray band in (a) marks the range of (c) and (d). (c), (d) Far-field polarization around $|U\rangle$ across the Janus window near $\rho_y=0.92$, calculated (c) from the model and (d) by RCWA. Colors show the polarization orientation angle $\psi$, and green lines mark $|E_x^{\sigma}|=|E_y^{\sigma}|$.}
\end{figure}

The same three kinds of flips occur for $|L\rangle$, for which Eq.~\eqref{eq:cd2d} holds with $\lambda_L$ in place of $\lambda_U$ and with $\cos\xi$ and $\sin\xi$ replaced by $-\sin\xi$ and $\cos\xi$. Figures~\ref{fig:2dq}(a) and \ref{fig:2dq}(b) show the channel charges of $|U\rangle$ and $|L\rangle$ for $\rho_x=0.55$. The flip of $|U\rangle$ near $\rho_y=0.453$ and those of $|L\rangle$ near $0.432$, $0.492$, $0.742$, and $0.771$ lie close to the crossings with the leaky states in Fig.~\ref{fig:2d}(c) and are of the first two kinds, whereas the others occur away from any crossing and are of the third kind. Because the lattice is vertically asymmetric, the two channels flip at different values of $\rho_y$ at every flip, so that each flip opens a Janus window. At the crossings, the windows are narrower than $3\times10^{-6}$ in $\rho_y$, because both flips lie close to the crossing. Away from the crossings, they are $2.7\times10^{-4}$ to $3.5\times10^{-3}$ wide, near $\rho_y=0.333$, $0.515$, and $0.92$ for $|U\rangle$ and near $0.371$ and $0.514$ for $|L\rangle$. The window near $0.92$ lies between $\rho_y=0.918$ and $0.922$ in both the model and RCWA, and the window near $0.515$ lies $0.003$ higher in RCWA than in the model. Figures~\ref{fig:2dq}(c) and \ref{fig:2dq}(d) show the far-field polarization across the window near $\rho_y=0.92$ in decreasing $\rho_y$. Because $|d^\sigma/c^\sigma|$ exceeds 100 there, $\psi$ rotates within a narrow strip around the $k_y$ axis, and the maps are drawn with $k_x$ magnified 100 times relative to $k_y$. In each channel, four C points lie on the lines $|E_x^\sigma|=|E_y^\sigma|$ before the flip of that channel and merge with the $\Gamma$-point BIC at the flip, which reverses the charge of that channel from $-1$ to $+1$, first in the downward channel and then in the upward channel.

\subsection{Square pillars}
For square pillars, $\rho_x=\rho_y$, the lattice has $C_{4v}$ symmetry, and the charge cannot flip. To see this for any modulation strength, we use the circular components $k_\pm=k_x\pm ik_y$ and $E_\pm^\sigma=E_x^\sigma\pm iE_y^\sigma$, for which $E_+^\sigma E_-^{\sigma*}=S_1+iS_2$ and $\psi=\frac12\arg\big(E_+^\sigma E_-^{\sigma*}\big)$. To first order in $\mathbf k$, the far field near a nondegenerate $\Gamma$-point BIC is
\begin{equation}\label{eq:Epm}
E_+^\sigma=\mathcal A^\sigma k_++\mathcal B^\sigma k_-,
\end{equation}
with the co-rotating coefficient $\mathcal A^\sigma$ and the counter-rotating coefficient $\mathcal B^\sigma$. The mirror $x\to-x$ maps $k_+\to-k_-$ and $E_+\to-E_-$, so that $E_-^\sigma=\pm(\mathcal B^\sigma k_++\mathcal A^\sigma k_-)$. Because $\psi=\frac12\arg\big(E_+^\sigma E_-^{\sigma*}\big)$, the charge is half the difference of the winding numbers of $E_+^\sigma$ and $E_-^\sigma$ around the $\Gamma$ point. A linear form $ak_++bk_-$ winds once around zero, positively for $|a|>|b|$ and negatively for $|a|<|b|$, so that $E_+^\sigma$ and $E_-^\sigma$ wind in opposite senses and
\begin{equation}\label{eq:qAB}
q^\sigma=\sgn\big(|\mathcal A^\sigma|-|\mathcal B^\sigma|\big).
\end{equation}
For $\mathbf E_{\mathrm{rad}}^\sigma=(c^\sigma k_y,\,d^\sigma k_x)$, $\mathcal A^\sigma=\frac i2(d^\sigma-c^\sigma)$ and $\mathcal B^\sigma=\frac i2(d^\sigma+c^\sigma)$, so that $|\mathcal A^\sigma|^2-|\mathcal B^\sigma|^2=-\Rea(c^{\sigma*}d^\sigma)$, and Eq.~\eqref{eq:qAB} reproduces Eq.~\eqref{eq:qcd}.

A rotation by an angle $\alpha$ about $z$ multiplies $k_\pm$ and $E_\pm^\sigma$ by $e^{\pm i\alpha}$, and a nondegenerate BIC transforms with a character $\chi=\pm1$, so that $E_+^\sigma(R\mathbf k)=\chi e^{i\alpha}E_+^\sigma(\mathbf k)$. Inserting Eq.~\eqref{eq:Epm} gives
\begin{equation}
\mathcal A^\sigma\big(1-\chi\big)=0,\qquad \mathcal B^\sigma\big(e^{-2i\alpha}-\chi\big)=0.
\end{equation}
For the twofold rotation of $C_{2v}$, $\alpha=\pi$, and a BIC, which does not radiate at the $\Gamma$ point, has $\chi=1$. Both $\mathcal A^\sigma$ and $\mathcal B^\sigma$ are then allowed, and the charge flips where $|\mathcal A^\sigma|=|\mathcal B^\sigma|$, a single real condition that differs between the two channels in a vertically asymmetric lattice. For the fourfold rotation of $C_{4v}$, $\alpha=\pi/2$ and $e^{-2i\alpha}=-1$. A BIC with $\chi=1$ has $\mathcal B^\sigma=0$ and $q^\sigma=+1$, and a BIC with $\chi=-1$ has $\mathcal A^\sigma=0$ and $q^\sigma=-1$. The charge is thus fixed by the symmetry of the BIC, the same in both channels, and independent of the pillar size, the slab, and the claddings. A change would require the remaining coefficient to vanish, which happens where other polarization singularities merge with the BIC. On either side of such a point, the charge keeps the value fixed by the symmetry of the BIC. Apart from such points, square pillars support neither charge flips nor Janus BICs as long as the $C_{4v}$ symmetry is kept, and the rectangular shape, which lowers the symmetry to $C_{2v}$, is what allows both. Because this argument uses only symmetry, it holds for any modulation strength, including the 2D zero-contrast gratings (ZCGs) of Figs.~6(e) and 6(f) of the main text.

\section{S6. Numerical methods}
\emph{Model.} The profile $\phi$, the Bragg frequency $\Omega_0$, and the Green functions of the three-layer reference slab are computed analytically, and the overlap integrals numerically across the grating layer. The radiation factors, the charges, the far-field maps, and the C points of Figs.~2--5 of the main text are computed from the model itself rather than from the expansions near the $\Gamma$ point. At each $\mathbf k$, the guided waves and the intermediate harmonics are placed at their own in-plane wave vectors, the coupled-mode matrix of Eq.~\eqref{eq:KHgen} is rebuilt, and the frequency at which it is evaluated is iterated to the eigenfrequency of the BIC band. The radiated field of each channel is Eq.~\eqref{eq:Esig} with the source of this eigenvector, where the TE form of the radiation is used for both in-plane components, which is exact on the $k_x$ and $k_y$ axes and differs from the TM form by $O(|\mathbf k|^2)$ off the axes. The factors in Eqs.~(9) and (10) of the main text are the values of this calculation at the $\Gamma$-point BIC, with $F_y^\sigma$ obtained from $E_y^\sigma/(A+B)$ through Eq.~\eqref{eq:Fyk} at $\mathbf k=(10^{-9}K,0)$. The second stop band of Fig.~5 is treated in the same way. The 2D model of Sec.~S5 includes the harmonics $|m|,|n|\le7$ and is evaluated in the same way.

\emph{RCWA.} The 1D lattices are computed with Fourier orders $|m|\le30$, the Janus-window widths of Table~I of the main text with $|m|\le40$, and the charges near the band flips of Fig.~2 with $|m|\le50$, which are unchanged at $|m|\le70$. The 2D lattices of Sec.~S5 use orders up to 5 along $x$ and 4 along $y$, and the 2D ZCGs orders up to 9 in each direction, raised to 11 at five cells near the edges of the Janus regions. Each BIC is located from the landscape of the smallest singular value of the boundary matrix, accepted when the ratio of the smallest to the largest singular value is below $10^{-14}$, and then followed in $\mathbf k$ and in the lattice parameter by continuation. The far fields of the two channels are the zeroth-order amplitudes in the cover and the substrate.

\emph{Finite-element check.} The charge flips of Fig.~2, the Janus windows of Figs.~3--5, and the ZCGs of Fig.~6 of the main text are also computed by the finite-element method (FEM) with COMSOL Multiphysics 6.0 and 6.3 (eigenfrequency study), in a cell of one period with Floquet conditions and perfectly matched layers in the cover and the substrate. The mesh is made mirror symmetric about the center of the ridge or pillar, because an asymmetric mesh breaks the symmetry that protects the BIC. The far field of each channel is taken from the zeroth diffraction order, as in RCWA. The 2D ZCGs are computed in a three-dimensional cell of one period in $x$ and $y$.

\section{S7. Cross-validation}
\FloatBarrier

Tables~\ref{tab:fem2}--\ref{tab:fem6} cross-validate the RCWA results with FEM for Figs.~2--6 of the main text. For the lattices of Figs.~2--5, the BIC frequencies agree to within $10^{-7}$ and the flips to within $2\times10^{-6}$. The band flips of Figs.~4 and 5(e) differ between the two methods by amounts comparable to the window widths, so that the band-flip-induced windows are measured from the band flip of each method, and they agree to within 1.2\%. For the 1D ZCGs of Fig.~6, the flips agree to within $10^{-5}$ at the second stop band and to within $6\times10^{-4}$ at the fourth, comparable to the shifts of the FEM flips when the far-field evaluation line or the mesh is changed. For the 2D ZCGs of Figs.~6(e) and 6(f), the Janus regions calculated by FEM agree well with those of RCWA.

\begin{table}[h]
\caption{\label{tab:fem2}Charge flips of the SiO$_2$/SiO$_2$ lattice of Fig.~2 of the main text ($\epsilon_{\mathrm{avg}}=12$, $\Delta\epsilon=1$, $t=0.5\Lambda$, fourth stop band), calculated by RCWA and FEM. $\omega_{\mathrm B}$ is the BIC frequency $\omega\Lambda/2\pi c$ at the $\Gamma$ point.}
\begin{ruledtabular}
\begin{tabular}{lll}
Quantity & RCWA & FEM \\
\hline
$\omega_{\mathrm B}$ at $\rho=0.24196$ & $0.6164566$ & $0.6164567$ \\
$\rho_{\mathrm{F1}}$ (band flip) & $0.2419645$ & $0.2419644$ \\
$\rho_{\mathrm{F2}}$ (band flip) & $0.4897511$ & $0.4897508$ \\
$\rho_{\mathrm{F3}}$ (merging) & $0.523078$ & $0.5230776$ \\
$\rho_{\mathrm{F4}}$ (merging) & $0.6601105$ & $0.6601099$ \\
$\rho_{\mathrm{F5}}$ (band flip) & $0.7414863$ & $0.7414863$ \\
\end{tabular}
\end{ruledtabular}
\end{table}

\begin{table}[h]
\caption{\label{tab:fem3}Janus window of the air/SiO$_2$ lattice of Fig.~3 of the main text ($\epsilon_{\mathrm{avg}}=12$, $\Delta\epsilon=1$, $t=0.5\Lambda$, fourth stop band), calculated by RCWA and FEM. $\rho^{\sigma}$ is the flip of channel $\sigma$.}
\begin{ruledtabular}
\begin{tabular}{lll}
Quantity & RCWA & FEM \\
\hline
$\omega_{\mathrm B}$ at $\rho=0.663$ & $0.6187106$ & $0.6187107$ \\
$(q^{\uparrow},q^{\downarrow})$ in window & $(-1,+1)$ & $(-1,+1)$ \\
$\rho^{\downarrow}$ & $0.6609392$ & $0.6609377$ \\
$\rho^{\uparrow}$ & $0.6648944$ & $0.6648928$ \\
width & $3.955\times10^{-3}$ & $3.955\times10^{-3}$ \\
\end{tabular}
\end{ruledtabular}
\end{table}

\begin{table}[h]
\caption{\label{tab:fem4}Janus window at the first band flip of the lattice of Fig.~4 of the main text (air/SiO$_2$, $\epsilon_{\mathrm{avg}}=12$, $\Delta\epsilon=1$, $t=0.5\Lambda$, fourth stop band), calculated by RCWA and FEM. $\Delta\rho^{\sigma}=\rho^{\sigma}-\rho_{\mathrm{BF}}$ is measured from the band flip $\rho_{\mathrm{BF}}$ of each method.}
\begin{ruledtabular}
\begin{tabular}{lll}
Quantity & RCWA & FEM \\
\hline
$\omega_{\mathrm B}$ at $\rho=0.24173$ & $0.6170109$ & $0.6170110$ \\
$\rho_{\mathrm{BF}}$ & $0.2417328$ & $0.2417325$ \\
$(q^{\uparrow},q^{\downarrow})$ in window & $(+1,-1)$ & $(+1,-1)$ \\
$\Delta\rho^{\downarrow}$ & $4.27\times10^{-6}$ & $4.26\times10^{-6}$ \\
$\Delta\rho^{\uparrow}$ & $4.44\times10^{-6}$ & $4.44\times10^{-6}$ \\
width & $1.77\times10^{-7}$ & $1.75\times10^{-7}$ \\
\end{tabular}
\end{ruledtabular}
\end{table}

\begin{table}[h]
\caption{\label{tab:fem5e}Band-flip-induced Janus window of Fig.~5(e) of the main text (air/SiO$_2$, $\epsilon_{\mathrm{avg}}=12$, $\Delta\epsilon=1$, $t=0.5\Lambda$, second stop band), calculated by RCWA and FEM, with the quantities defined as in Table~\ref{tab:fem4}.}
\begin{ruledtabular}
\begin{tabular}{lll}
Quantity & RCWA & FEM \\
\hline
$\omega_{\mathrm B}$ at $\rho=0.48917$ & $0.33942387$ & $0.33942388$ \\
$\rho_{\mathrm{BF}}$ & $0.4891703$ & $0.4891698$ \\
$(q^{\uparrow},q^{\downarrow})$ in window & $(+1,-1)$ & $(+1,-1)$ \\
$\Delta\rho^{\downarrow}$ & $3.44\times10^{-7}$ & $3.40\times10^{-7}$ \\
$\Delta\rho^{\uparrow}$ & $4.26\times10^{-7}$ & $4.23\times10^{-7}$ \\
width & $8.22\times10^{-8}$ & $8.21\times10^{-8}$ \\
\end{tabular}
\end{ruledtabular}
\end{table}

\begin{table}[h]
\caption{\label{tab:fem5f}Merging-induced Janus windows of Fig.~5(f) of the main text (air/SiO$_2$, $\epsilon_{\mathrm{avg}}=12$, $\Delta\epsilon=1$, $\rho=0.30$, second stop band), calculated by RCWA and FEM. $t^{\sigma}$ is the flip of channel $\sigma$.}
\begin{ruledtabular}
\begin{tabular}{lll}
Quantity & RCWA & FEM \\
\hline
$\omega_{\mathrm B}$ at $t=1.159\Lambda$ & $0.30640655$ & $0.30640655$ \\
$(q^{\uparrow},q^{\downarrow})$ in window 1 & $(+1,-1)$ & $(+1,-1)$ \\
$t^{\downarrow}$ & $1.155651\Lambda$ & $1.155650\Lambda$ \\
$t^{\uparrow}$ & $1.162453\Lambda$ & $1.162453\Lambda$ \\
width & $6.802\times10^{-3}\Lambda$ & $6.803\times10^{-3}\Lambda$ \\
\hline
$(q^{\uparrow},q^{\downarrow})$ in window 2 & $(-1,+1)$ & $(-1,+1)$ \\
$t^{\downarrow}$ & $1.162837\Lambda$ & $1.162839\Lambda$ \\
$t^{\uparrow}$ & $1.168429\Lambda$ & $1.168429\Lambda$ \\
width & $5.592\times10^{-3}\Lambda$ & $5.590\times10^{-3}\Lambda$ \\
\end{tabular}
\end{ruledtabular}
\end{table}

\begin{table}[h]
\caption{\label{tab:fem6}Janus windows of the 1D Si ZCGs of Fig.~6 of the main text ($\epsilon_{\mathrm{Si}}=12.1104$, $\epsilon_{\mathrm{SiO_2}}=2.1316$, air cover), at the second stop band in Fig.~6(b) ($t=0.1\Lambda$, $h=0.4\Lambda$, $\omega_{\mathrm B}$ at $\rho=0.35$) and the fourth stop band in Fig.~6(d) ($t=0.45\Lambda$, $h=0.2\Lambda$, $\omega_{\mathrm B}$ at $\rho=0.30$), calculated by RCWA and FEM.}
\begin{ruledtabular}
\begin{tabular}{lllll}
 & \multicolumn{2}{c}{Fig.~6(b)} & \multicolumn{2}{c}{Fig.~6(d)} \\
Quantity & RCWA & FEM & RCWA & FEM \\
\hline
$\omega_{\mathrm B}$ & $0.4636748$ & $0.4636737$ & $0.6159811$ & $0.6159812$ \\
$(q^{\uparrow},q^{\downarrow})$ in window & $(+1,-1)$ & $(+1,-1)$ & $(+1,-1)$ & $(+1,-1)$ \\
$\rho^{\downarrow}$ & $0.350478$ & $0.350483$ & $0.30147$ & $0.30175$ \\
$\rho^{\uparrow}$ & $0.412545$ & $0.412537$ & $0.48066$ & $0.48119$ \\
width & $6.207\times10^{-2}$ & $6.205\times10^{-2}$ & $0.1792$ & $0.1794$ \\
\end{tabular}
\end{ruledtabular}
\end{table}